\documentclass[longauth]{aa}

\usepackage{graphicx}
\usepackage{mwe}
\usepackage{longtable}
\usepackage{siunitx}
\usepackage[table]{xcolor}
\usepackage{textcomp}
\usepackage{chemist}
\usepackage{multirow}
\usepackage{caption}
\usepackage{subcaption}
\usepackage{float}
\usepackage{txfonts}

\titlerunning{}

\usepackage{natbib,twoopt}

\usepackage[colorlinks=true, pdfstartview=FitV, linkcolor=blue, citecolor=blue, urlcolor=blue, breaklinks=true]{hyperref} 
\bibpunct{(}{)}{;}{a}{}{,}             
\makeatletter
  \newcommandtwoopt{\citeads}[3][][]{\href{http://adsabs.harvard.edu/abs/#3}%
    {\def\hyper@linkstart##1##2{}%
     \let\hyper@linkend\@empty\citealp[#1][#2]{#3}}}
  \newcommandtwoopt{\citepads}[3][][]{\href{http://adsabs.harvard.edu/abs/#3}%
    {\def\hyper@linkstart##1##2{}%
     \let\hyper@linkend\@empty\citep[#1][#2]{#3}}}
  \newcommandtwoopt{\citetads}[3][][]{\href{http://adsabs.harvard.edu/abs/#3}%
    {\def\hyper@linkstart##1##2{}%
     \let\hyper@linkend\@empty\citet[#1][#2]{#3}}}
  \newcommandtwoopt{\citeyearads}[3][][]%
    {\href{http://adsabs.harvard.edu/abs/#3}
    {\def\hyper@linkstart##1##2{}%
     \let\hyper@linkend\@empty\citeyear[#1][#2]{#3}}}
\makeatother

\usepackage{color}

\usepackage{color}

\usepackage{cleveref}

\newcommand{\Msolar}{\mbox{$\mathrm{M}_{\odot}\,$}}

\usepackage{chemist}
\begin{document}
\title{Mid-infrared circumstellar emission of the long-period Cepheid $\ell$~Carinae resolved with VLTI/MATISSE\thanks{Based on observations made with ESO telescopes at Paranal observatory  under  program  ID 0104.D-0554(A)} }

\titlerunning{Mid-infrared circumstellar emission of the long-period Cepheid $\ell$~Carinae resolved with VLTI/MATISSE}
\authorrunning{Hocd\'e et al. }
\author{V. Hocd\'e \inst{1} 
\and N. Nardetto \inst{1}
\and A. Matter \inst{1}
\and E. Lagadec  \inst{1}
\and A. M\'erand \inst{2}
\and P. Cruzal\`ebes \inst{1}
\and A. Meilland \inst{1}
\and F. Millour \inst{1}
\and B. Lopez \inst{1}
\and P.~Berio \inst{1}
\and G. Weigelt \inst{3}
% ...
}

\institute{Universit\'e Côte d'Azur, Observatoire de la C\^ote d'Azur, CNRS, Laboratoire Lagrange, France,\\
email : \texttt{vincent.hocde@oca.eu}
\and European Southern Observatory, Karl-Schwarzschild-Str. 2, 85748 Garching, Germany
\and Max-Planck-Institut f\"ur Radioastronomie, Auf dem H\"ugel 69, D-53121 Bonn, Germany
% ...
}

\date{Received ... ; accepted ...}

\abstract{The nature of circumstellar envelopes (CSE) around Cepheids is still a matter of debate. The physical origin of their infrared (IR) excess could be either a shell of ionized gas, or a dust envelope, or both.}{This study aims at constraining the geometry and the IR excess of the environment of the bright long-period Cepheid $\ell$~Car (P=35.5 days) at mid-IR wavelengths to understand its physical nature.}{
We first use photometric observations in various bands (from the visible domain to the infrared) and \textit{Spitzer} Space Telescope spectroscopy to constrain the IR excess of $\ell$ Car. Then, we analyze the VLTI/MATISSE measurements at a specific phase of observation, in order to determine the flux contribution, the size and shape of the environment of the star in the $L$ band. We finally test the hypothesis of a shell of ionized gas in order to model the IR excess.} {We report the first detection in the $L$ band of a centro-symmetric extended emission around $\ell$ Car, of about $1.7\,R_\star$ in full width at half maximum, producing an excess of about $7.0\%$ in this band. This latter value is used to calibrate the IR excess found when comparing the photometric observations in various bands and quasi-static atmosphere models. In the $N$ band, there is no clear evidence for dust emission from VLTI/MATISSE correlated flux and \textit{Spitzer} data. On the other side, the modeled shell of ionized gas implies a more compact CSE ($1.13\pm0.02\,R_\star$) and fainter (IR excess of 1\% in the $L$ band).}
{We provide new evidences for a compact CSE of $\ell$ Car and we demonstrate the capabilities of VLTI/MATISSE for determining common properties of CSEs. While the compact CSE of $\ell$ Car is probably of gaseous nature, the tested model of a shell of ionized gas is not able to simultaneously reproduce the IR excess and the interferometric observations. Further Galactic Cepheids observations with VLTI/MATISSE are necessary for determining the properties of CSEs, which may also depend on both the pulsation period and the evolutionary state of the stars.}

\keywords{Techniques : Interferometry -- Infrared : CSE -- stars: variables: Cepheids – stars: atmospheres}
\maketitle

\section{Introduction}\label{s_Introduction}
Circumstellar envelopes (CSEs) around Cepheids have been spatially resolved by long-baseline interferometry in the $K$ band with the Very Large Telescope Interferometer (VLTI) and the Center for High Angular Resolution Astronomy (CHARA) \citep{kervella06a,merand06}. They were detected around four Cepheids including $\ell$ Car, in the $N$~band with VLTI/VISIR and VLTI/MIDI \citep{kervella09,gallenne13b}. In the $K$ band, the diameter of the envelope appears to be at least 2 stellar radii and the flux contribution is up to 5\% of the continuum \citep{merand07}. Both the size and flux contribution are expected to be larger in the $N$ band, as demonstrated by \cite{kervella09} for $\ell$ Car, \cite{gallenne13b} for X~Sgr and T~Mon, and by Gallenne et al. (2021, submitted) for other Cepheids. Moreover, a CSE has also been discovered in the visible domain with the CHARA/VEGA instrument around $\delta$~Cep \citep{nardetto16}. The systematic presence of a CSE is still a matter of debate. While some studies have found no significant observational evidence for circumstellar dust envelopes in a large number of Cepheids \citep{Schmidt2015,Gro2020}, Gallenne et al. (2021, submitted) found a significant IR excess attributed to a CSE for 10 out of 45 Cepheids.\\
\indent Recently, \cite{Hocde2020a} (hereafter Paper I) used an analytical model of free-free and bound-free emission from a thin shell of ionized gas to explain the near and mid-IR excess of Cepheids. They found a typical radius for this shell of ionized gas of about $R_\mathrm{shell}$=1.15$\,R_\star$. This shell of ionized gas could be due to periodic shocks occurring in both the atmosphere and the chromosphere, which heat up and ionize the gas. Using VLT/UVES data, \citep{Hocde2020b} found a radius for the chromosphere of at least $R_\mathrm{chromo}$=1.5$\,R_\star$ in long-period Cepheids. In addition, they found a motionless H$\alpha$ feature in UVES high-resolution spectra obtained for several long-period Cepheids, including $\ell$~Car. This absorption feature was attributed to a static CSE surrounding the chromosphere above at least 1.5$\,R_\star$, and reported by various authors around $\ell$ Car \citep{Rodgers1968,love1994,nardetto08b}.\\
\indent Determining the IR excess and the size of the CSE is a key to understand the physical processes at play. In this paper, we study the long-period Cepheid $\ell$ Car. Our aim is to determine its IR excess from photometric measurements in various bands and \textit{Spitzer} spectroscopy, while inferring the size of the CSE and its flux contribution thanks to the unique capabilities provided by the Multi AperTure mid-Infrared SpectroScopic Experiment (VLTI/MATISSE; \cite{Lopez2014,Allouche2016,Robbe2018}) in the $L$ (2.8-4.0$\,\mu$m), $M$ (4.5 - 5$\,\mu$m) and $N$ bands (8 - 13$\,\mu$m). \\
\indent We first reconstruct the IR excess in Sect \ref{sect2:obs}. Then, we present the data reduction and calibration process of VLTI/MATISSE data in Sect.~\ref{sect:matisse_obs}. We use a simple model of the CSE  to reproduce the visibility measurement of VLTI/MATISSE in  Sect.~\ref{sect:envelope_model}. In Sect.~\ref{sect:discussion} we discuss the physical origin of this envelope, and present a model of a shell of ionized gas.  We summarize our conclusions in Sect.~\ref{sect:conclusion}.

\section{Deriving the infrared excess of $\ell$ Car}\label{sect2:obs}

\subsection{Near-IR excess modeling with \texttt{SPIPS}}\label{sect:spips_photometry}
Due to their intrisic variability, the photospheres of the Cepheids are difficult to model along the pulsation cycle. However, it is an essential prerequisite for deriving both the IR excess in a given photometric band, and the expected angular stellar diameter at a specific phase of interferometric observations (required in Sect.~\ref{sect:matisse_obs}). In order to model the photosphere we use SpectroPhoto-Interferometric modeling of Pulsating Stars (\texttt{SPIPS}), which is a model-based parallax-of-pulsation code which includes photometric, interferometric, effective temperature and radial velocity measurements in a robust model fitting process \citep{merand15}. 

\begin{figure}[]
    \centering
    
    \begin{subfigure}[b]{0.52\textwidth}
        \centering
        \includegraphics[width=\textwidth]{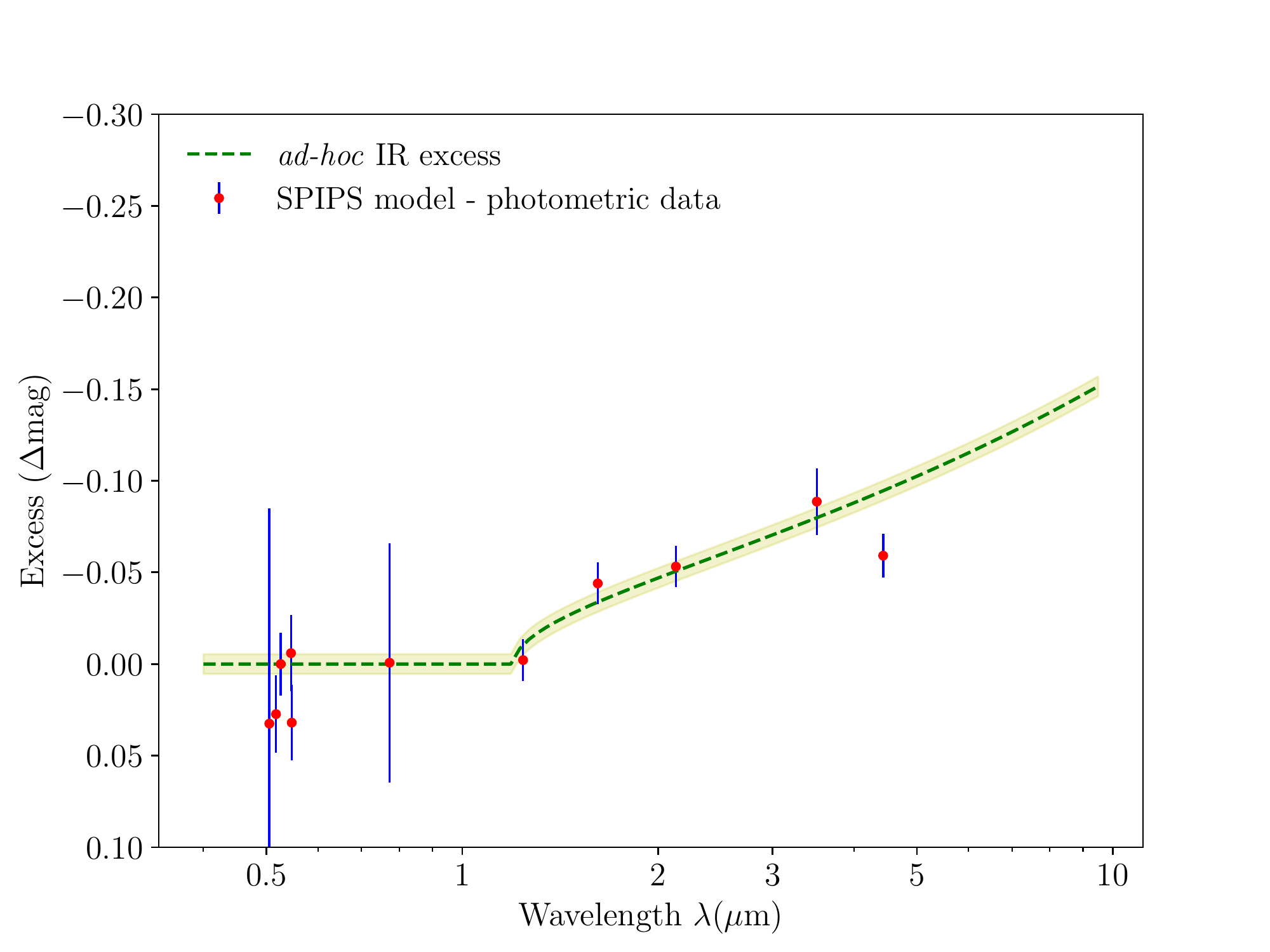}
        \caption[Network2]%
            {\small  \texttt{SPIPS} cycle-averaged IR excess \label{fig:excess-spips}}  
            \vspace{-0.3cm}
        \end{subfigure}
        \vskip\baselineskip
        
        \begin{subfigure}[b]{0.52\textwidth}   
            \centering 
            \includegraphics[width=\textwidth]{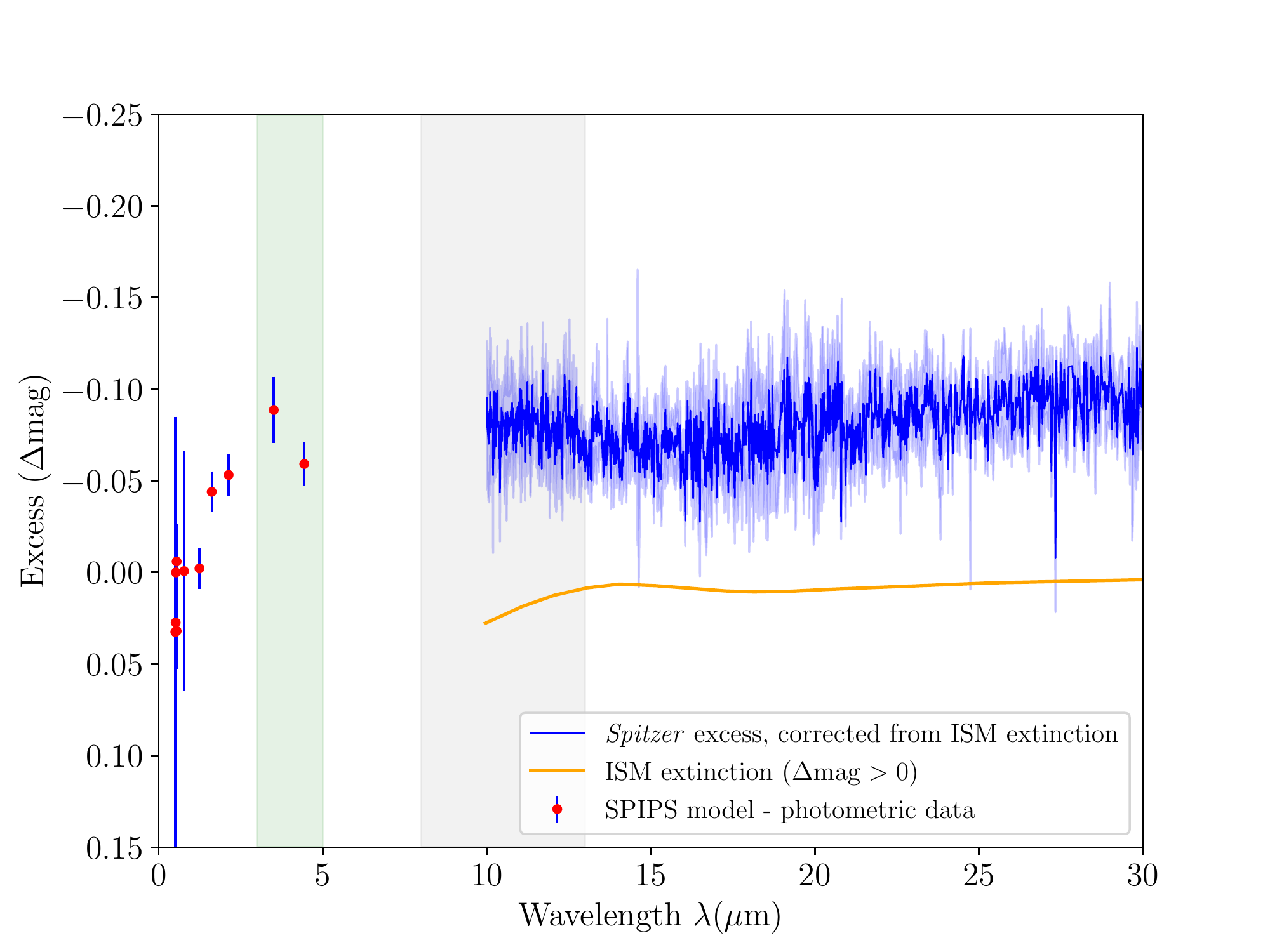}
            \caption[]%
            {\small Reconstructed IR excess with \texttt{SPIPS} and \textit{Spitzer}}  \label{fig:ir_excess}
                     \vspace{-0.3cm}
        \end{subfigure}
        
        \caption[]
        {\small (a) Averaged IR excess of $\ell$ Car as derived with the \texttt{SPIPS} algorithm (red dots) and presented with the fitted {\it ad-hoc} analytic law $\mathrm{IR}_\mathrm{ex}=-0.053(\lambda-1.2)^{0.501}\mathrm{mag}$. For each photometric band (described in Appendix \ref{app:spips}), the red dots are the mean excess value averaged over the cycle of pulsation of $\ell$ Car while the blue uncertainties correspond to the respective standard deviation. (b) The IR excess of $\ell$ Car is reconstructed at the specific phase of \textit{Spitzer} observation. It includes the averaged IR excess from \texttt{SPIPS} (red dots) and the \textit{Spitzer} observations (blue line) corrected for the silicate absorption of the ISM along the line of sight (orange curve). The silicate extinction has been derived using silicate refractive index from \cite{DL}. Green and gray vertical stripes represent the $LM$ and $N$ bands of MATISSE, respectively.}
    \end{figure}

\texttt{SPIPS} uses a grid of ATLAS9 atmospheric models\footnote{\url{http://wwwuser.oats.inaf.it/castelli/grids.html}} \citep{castelli2003} with solar metallicity and a standard turbulent velocity of 2 km/s. \texttt{SPIPS} was already extensively described and used in several studies \citep{merand15,breitfelder16,kervella17,Gallenne2017}, Javanmardi et al. (2021, accepted), Gallenne et al. (2021, submitted), and Paper~I. \texttt{SPIPS} also takes into account the possible presence of an IR excess by fitting an {\it ad-hoc} analytic power law $\mathrm{IR}_\mathrm{ex}$ (see green curve in Fig.~\ref{fig:excess-spips}). We note that Gallenne et al. (2021, submitted) have shown that $\ell$ Car possibly has an IR excess using \texttt{SPIPS}, but its low detection level (2.9 ± 3.2\% in $K$ band) is not significant compared to a single-star model (i.e. without IR excess model). \cite{breitfelder16} also provided a \texttt{SPIPS} fitting of $\ell$~Car which did not present any significant IR excess. However, this work was done by considering only the $V$, $J$, $H$ and $K$ photometric bands, i.e. not the mid-infrared bands, which are critical to detect such an IR excess. Finally, since a CSE around $\ell$ Car has been detected by interferometry in the infrared \citep{kervella06a,kervella09}, we enable \texttt{SPIPS} to fit an IR excess in the following analysis.
\texttt{SPIPS} provides a fit to the photometry along the pulsation cycle, which is in agreement with the observational data (description in Appendix \ref{l_car_spips}). From Fig.~\ref{l_car_spips}, we emphasize that the \texttt{SPIPS} fitted model is well constrained by the numerous observations, and thus the physical parameters of the photosphere are accurately derived. However, as noted in the \texttt{SPIPS} original paper, the uncertainties on the derived parameters are purely statistical and do not take into account systematics. While the fitting is satisfactory from the visible domain to the near infrared, \texttt{SPIPS} fits an IR excess for wavelengths above about 1.2$\,\mu$m. Indeed the observed brightness ($m_\mathrm{obs}$) significantly exceeds the predictions ($m_\mathrm{kurucz}$). The averaged IR excess over a pulsation cycle is presented in Fig.~\ref{fig:excess-spips}. In the next section we combine this average IR excess with \textit{Spitzer} mid-infrared spectroscopy.

\subsection{Mid-IR excess from \textit{Spitzer} observations}\label{sect:spitzer}

Space-based observations such as those of obtained with the \textit{Spitzer} space telescope are useful to avoid perturbation caused by the Earth atmosphere, which is essential for the study of dust spectral features. We use spectroscopic observations made with the InfraRed Spectrograph IRS \citep{2004ApJS..154...18H} on board the \textit{Spitzer} telescope \citep{2004ApJS..154....1W}.  High spectral resolution (R=600) from 10$\,\mu$m to 38$\,\mu$m was used with Short-High (SH) and Long-High (LH) modules placed in the focal plane instrument. The full spectra were retrieved from the CASSIS atlas \citep{CASSIS} and the best flux calibrated spectrum was obtained from the optimal extraction using differential method, which eliminates low-level rogue pixels \citep{Lebouteiller2015}. Table~\ref{Tab.cals} provides an overview of the \textit{Spitzer} observation. The reference epoch at maximum light (MJD$_0$=50583.742), pulsation  period  ($P$=35.557$\,$days) and the rate of period change ($\dot{P}$=33.4$\pm1.0\,$s/yr) used to derive the pulsation phase corresponding to the \textit{Spitzer} observation are those computed by the \texttt{SPIPS} fitting of $\ell$~Car. We note that the most up-to-date and independent value of rate of period change, derived by \cite{Neilson2016} ($\dot{P}=20.23\pm1.38\,$s/y) is slightly lower than the value estimated by SPIPS. However adopting this latter value would not change the result of this paper.

\begin{table*}[tbp]
\caption{\label{Tab.cals} {\small \textit{Spitzer} data set of $\ell$ Car (program number 40968). The Astronomical Observation Request (AOR), the  date of observation,  the corresponding Modified Julian Date (MJD), and the  pulsation phase ($\phi$) are indicated. The physical parameters of $\ell$ Car derived by \texttt{SPIPS} at the \textit{Spitzer} observation phase are indicated: the effective temperature ($T_\mathrm{eff}$), the surface gravity (log $g$) and the limb-darkened angular diameter ($\theta_\mathrm{LD}$). Statistical uncertainties on $T_\mathrm{eff}$ and $\theta$ are provided by \texttt{SPIPS}, whereas uncertainty on log $g$ is arbitrarily set to 10\%.}}
\begin{center}
\begin{tabular}{ccccc|ccc}
\hline
\hline
Object  &       AOR    &       Date & MJD & $\phi_\textit{Spitzer} $  &  $T_\mathrm{eff}(\phi_\textit{Spitzer})$ &  log $g(\phi_\textit{Spitzer})$& $\theta_\mathrm{LD}(\phi_\textit{Spitzer}) $ \\
  &           &  (yyyy.mm.dd)      & (days) &   &  (K) &  (cgs) & (mas) \\
\hline                                                                  
$\ell$ Car &      	23403520         &               2008.05.03&     54589.502 & 0.648 & $4630^{+15}_{-15}$& $0.98^{+0.10}_{-0.10}$&$2.897^{+0.002}_{-0.002}$   \\[0.15cm]
\hline                                                                                                                                          \end{tabular}
\normalsize
\end{center}
\end{table*}

In order to derive the IR excess of \textit{Spitzer} observation, and correct for the absorption of the interstellar matter (ISM), we followed the method developed in Paper I. First, we derived the IR excess of $\ell$~Car at the specific \textit{Spitzer} epoch using
\begin{equation}\label{eq:mag_spitzer}
\Delta \mathrm{mag} = m_\textit{Spitzer} - m_\mathrm{kurucz}[\phi_\textit{Spitzer}]
\end{equation}

\noindent where  $m_\textit{Spitzer}$ is the magnitude of the \textit{Spitzer} observation and $m_\mathrm{kurucz}[\phi_\textit{Spitzer}]$ is the magnitude of the ATLAS9 atmospheric model interpolated at the phase of \textit{Spitzer} observations ($\phi_\textit{Spitzer}$) using the parameters derived by SPIPS and given in Table~\ref{Tab.cals}. We discarded wavelengths of \textit{Spitzer} observations longer than 30$\,\mu$m due to extremely large uncertainties. Secondly, we corrected the spectrum for ISM extinction by subtracting a synthetic ISM composed of silicates from \cite{DL} (see orange curve in Fig. \ref{fig:ir_excess}). This calculation assumes a relation between the extinction E($B-V$) derived by \texttt{SPIPS} (i.e. E($B-V$)=0.148 mag) and the silicate absorption from diffuse interstellar medium (see Eq.~4 in Paper~I). Finally, we combine this result with \texttt{SPIPS} averaged IR excess in order to reconstruct the IR excess from the visible to the mid-infrared domain. The determined IR excess is presented in Fig.~\ref{fig:ir_excess} (red dots and blue line).\\
\indent Similarly to the five Cepheids presented in Paper I, we observe a continuum IR excess increasing up to $-$0.1 mag at 10$\,\mu$m. In particular within the specific bands of MATISSE (see vertical strips in Fig. \ref{fig:ir_excess}), the IR excess is between $-$0.05 and $-$0.1$\,$mag at 3.5$\,\mu$m and 4.5$\,\mu$m ($L$ and $M$ bands) which represents an excess of 5 to 10\% above the stellar continuum. In addition, we find no silicate emission feature in the \textit{Spitzer} spectrum around 10$\,\mu$m ($N$ band) within the uncertainties, which points toward an absence of significant amount of circumstellar dust with amorphous silicate components.

\section{VLTI/MATISSE interferometric observations}\label{sect:matisse_obs}
MATISSE is the four-telescope beam combiner in the $L$, $M$ and $N$ bands of the Very Large Telescope Interferometer (VLTI). The VLTI array consists of four 1.8-meter auxiliary telescopes (ATs) and four 8-meter unit telescopes (UTs), and provides baseline lengths from 11 meters to 150 meters. 
As a spectro-interferometer, MATISSE provides dispersed fringes and photometries. The standard observing mode of MATISSE (so-called hybrid) uses the two following photometric measurement modes: SIPHOT for $L$ and $M$ bands, in which the photometry is measured simultaneously with the dispersed interference fringes, and HIGH SENS for $N$ band, in which the photometric flux is measured separately after the interferometric observations. The observations were carried out during the nights of 27 and 28 February 2020 with the so-called large configuration of the ATs quadruplet (A0-G1-J2-J3, with ground baseline lengths from 58 to 132$\,$m) in low spectral resolution (R=$\lambda /\Delta \lambda \approx $30). The log of the MATISSE observations  is  given  in  Table  \ref{Tab.log}. The raw data were processed using the version (1.5.5) of the MATISSE data reduction software \footnote{The MATISSE reduction pipeline is publicly available at \url{http://www.eso.org/sci/software/pipelines/matisse/.}}. The steps of the data reduction process are described in \citet{Millour2016}. 

The MATISSE absolute visibilities are estimated by dividing the measured correlated flux by the photometric flux. Thermal background effects affect the measurement in the $M$ and $N$ bands significantly more than in the $L$ band. Thus, we use only the $L$ band visibilities in our modeling and analysis,  while the $L$ and $M$ photometries are used for deriving the observed total flux (see Sect.~\ref{sect:flux_cal}). In the $N$ band, the total flux of $\ell$ Car ($\approx$17$\,$Jy) is at the lower limit of the ATs sensitivity with MATISSE for accurate visibility measurements ($\sim$~20$\,$Jy, as stated on the ESO webpage of the MATISSE instrument). Thus, we rather use the correlated flux measurements (not normalized by the photometric flux) as an estimate of the $N$ band photometry of $\ell$~Car. In the following of the paper, we discarded the 4.1 to 4.5$\,\mu$m spectral region, where the atmosphere is not transmissive, and also the noisy edges of the atmosphere spectral bands. As a result, we analyze the following spectral region in $L$ (3.1-3.75$\mu$m) for absolute visibility and flux, $M$~(4.75-4.9$\mu$m) for flux, and in $N$ (8.2-12$\mu$m) for correlated flux.
\subsection{Calibration of the squared visibility in the $L$ band}\label{sect:cal_sci}
\begin{figure}[h]
\begin{center}
\includegraphics[width=0.53\textwidth]{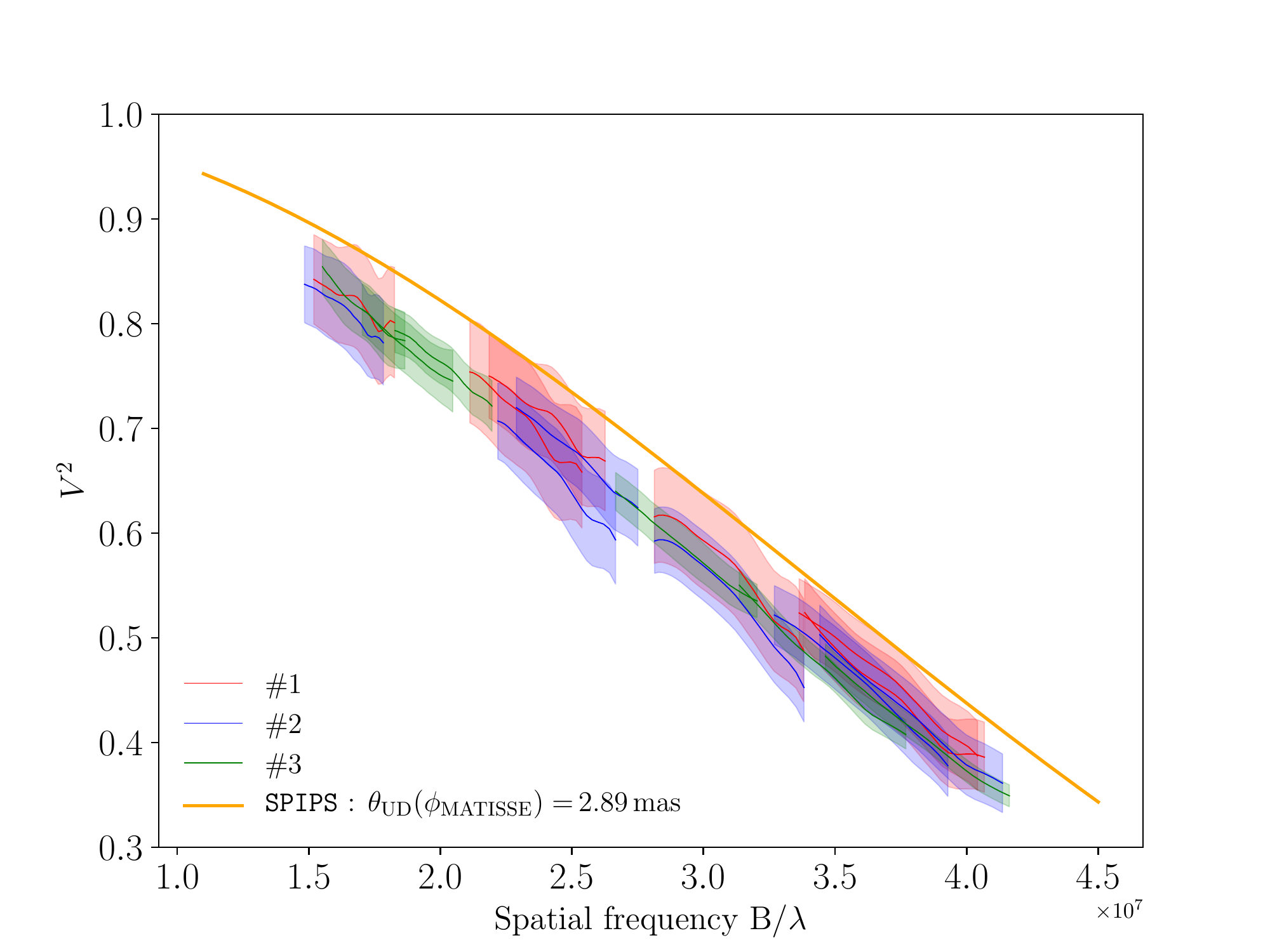}
\caption{\small The calibrated squared visibilities of $\ell$ Car as a function of the spatial frequencies in the $L$ band for observations $\#1$, 2 and 3. The theoretical visibility curve corresponding to a uniform disk of $\theta_\mathrm{UD}=2.887\pm0.003\,$mas in the $L$ band, as derived from the \texttt{SPIPS} analysis (i.e. without CSE) is indicated for comparison.  \label{fig:cal_sci}}  
 
\end{center}
\end{figure}

Calibrators were selected using the SearchCal tool of the JMMC\footnote{SearchCal is publicly available at \url{https://www.jmmc.fr/english/tools/proposal-preparation/search-cal/}}, with a high confidence level on the derived angular diameter ($\chi^2\leq5$, see Appendix A.2 in \cite{Chelli2016}). Moreover, calibrator fluxes in the $L$ and $N$ bands have to be of the same order than for the science target (i.e. $\approx$10 to 100$\,$Jy in $L$ and $N$) to ensure a reliable calibration \citep{cruz2019}. These flux requirements induce the use of partially resolved calibrators at the 132-m longest baseline. The $L$ band uniform-disk (UD) angular diameters for the standard stars as well as the corresponding $L$ and $N$ band fluxes and the spectral type are given in Table \ref{Tab.cal}.\\
\indent We analyze these calibration stars with a special care. The first calibrator of this night, q~Car, was observed in the frame of a different program and was not used in the interferometric visibility calibration process. Indeed, it is a supergiant star of spectral type KII, which makes its angular diameter of 5.2$\,$mas too large (compared to $\ell$~Car), and rather uncertain in the $L$ band. However, q~Car appears to be a suitable total flux calibrator for $\ell$~Car in both $L$ and $M$ bands (see Sect.~\ref{sect:flux_cal}). Indeed, although q Car is an irregular long-period variable, its photometric variation in the visible is $\Delta$mag$\approx$0.1 \footnote{\url{http://www.sai.msu.su/groups/cluster/gcvs/gcvs/iii/iii.dat}} \citep[GCVS,][]{GCVS2017}, thus its flux variation in the Rayleigh-Jeans domain, i.e. in the $L$ and $M$ bands, only represents about 1\%. Moreover q~Car was observed at an air mass very close to the one of $\ell$~Car (see Table \ref{Tab.log}). In the analysis, we discarded the calibrator B~Cen since it has inconsistent visibilities as explained in Appendix \ref{sect:b_cen}. Thus, we used $\epsilon$~Ant to calibrate observation $\#$1. For observation $\#$2, we also used $\epsilon$ Ant. For observation $\#3$, we bracketed the science target with the standard CAL-SCI-CAL strategy calibration using $\beta$~Vol. The calibrated squared visibilities ($V^2$) in $L$ band are presented in Fig.~\ref{fig:cal_sci}. The visibility curve associated with the limb-darkened angular diameter of the star (without CSE), which was derived from the \texttt{SPIPS} analysis at the specific phase of VLTI/MATISSE ($\phi$=0.07), is shown for comparison.

\begin{table}[]
\caption{\label{Tab.log} \small Log of the observations for nights of 27 February (observations $\#$1 and 2) and 28 February 2020 (observation $\#$3) with the Modified Julian Date (MJD), the pulsation phase $\phi$, the seeing (in arcsecond), the air mass (AM), integrated water vapour (IWV) at 30 deg elevation (in millimeter) and the coherence time $\tau_0$ (in millisecond) are indicated. Stars used in the visibility calibration strategy are marked with an asterisk (see Sect.~\ref{sect:cal_sci}). q Car and $\beta$ Vol are used to calibrate the total flux (see Sect.~\ref{sect:flux_cal}), while e Cen is used to check the consistency of the other calibrators (see Appendix~\ref{sect:b_cen}). The calibrator B Cen is not used, see Appendix~\ref{sect:b_cen}}.
\begin{center}
\begin{tabular}{c|l|c|c|c|c|c}
\hline
\hline
\#& Target	&	Date 	&  $\phi$  &	seeing 	&	AM &$\tau_0$ 	\\
&	& (MJD)		&		&	($
^{\prime \prime}$)		&	& (ms)	\\
\hline											
&q Car	&	58907.11	&		&	0.78		&1.40	& 6.0	\\

1&$\ell$ Car*	&	58907.12	&	0.05	&	0.66 &	 1.32 & 5.3	\\

&B Cen	&	58907.13	&		&	0.85	&1.39		&6.1	\\

2&$\ell$ Car*	&	58907.14 	&	0.05	& 1.15	& 1.27	 &6.0	\\

&$\epsilon$ Ant*	&	58907.15	&		&		0.69		&	1.02   &6.9	\\
&e Cen	&58907.28		&		&		1.26		&	1.10 & 4.8	\\
\hline
&bet Vol*	&58907.99		&		&	0.79	&   1.51	&6.3		\\
3&$\ell$ Car*	&	58908.00 	&	0.07		&0.81	&	1.66  &6.3\\
&bet Vol*	&	58908.01	&		& 0.91		& 1.42		& 15.7		\\
\hline

\hline
\end{tabular}
\normalsize
\end{center}
\end{table}

\begin{table}[]
\caption{\label{Tab.cal} \small Calibrator properties. $\theta_\mathrm{UD}$ is the uniform disk (UD) angular diameter in $L$ band from the JSDC V2 catalogue \citep{bourges14}, $F_L$ and $F_N$ are the flux in the $L$ and $N$ bands from the Mid-infrared stellar Diameters and Fluxes compilation Catalogue (MDFC) \citep{cruz2019}.}
\begin{center}
\begin{tabular}{l|c|c|c|c}
\hline
\hline
Calibrator	&	$\theta_\mathrm{UD}$(mas)	&	$F_L$ (Jy)	&	$F_N$(Jy)	&	Sp. Type	 \\
		\hline											
q Car	&	5.19$\pm$0.60 	&299.5$\pm$50.6	&	45.8$\pm8.2$	& K2.5 II		\\
B Cen	&	2.54$\pm$0.28	&72.3$\pm11.8$	&	10.6$\pm1.0$	& K3 III		\\
$\epsilon$ Ant	&	2.86$\pm$0.30	&88.1$\pm1.2$		&12.2$\pm1.8$		& K3 III	\\
e Cen   & 2.97$\pm$0.29	&88.9$\pm10.9$		&13.2$\pm1.6$		& K3.5 III	\\
bet Vol	&	2.90$\pm$0.29	&95.4$\pm23.5$		&	14.6$\pm3.4$	& K2 III			\\
\hline

\hline
\end{tabular}
\normalsize
\end{center}
\end{table}

\begin{figure*}[htb]
    \centering 
\begin{subfigure}[b]{.5\textwidth}
  \includegraphics[width=\linewidth]{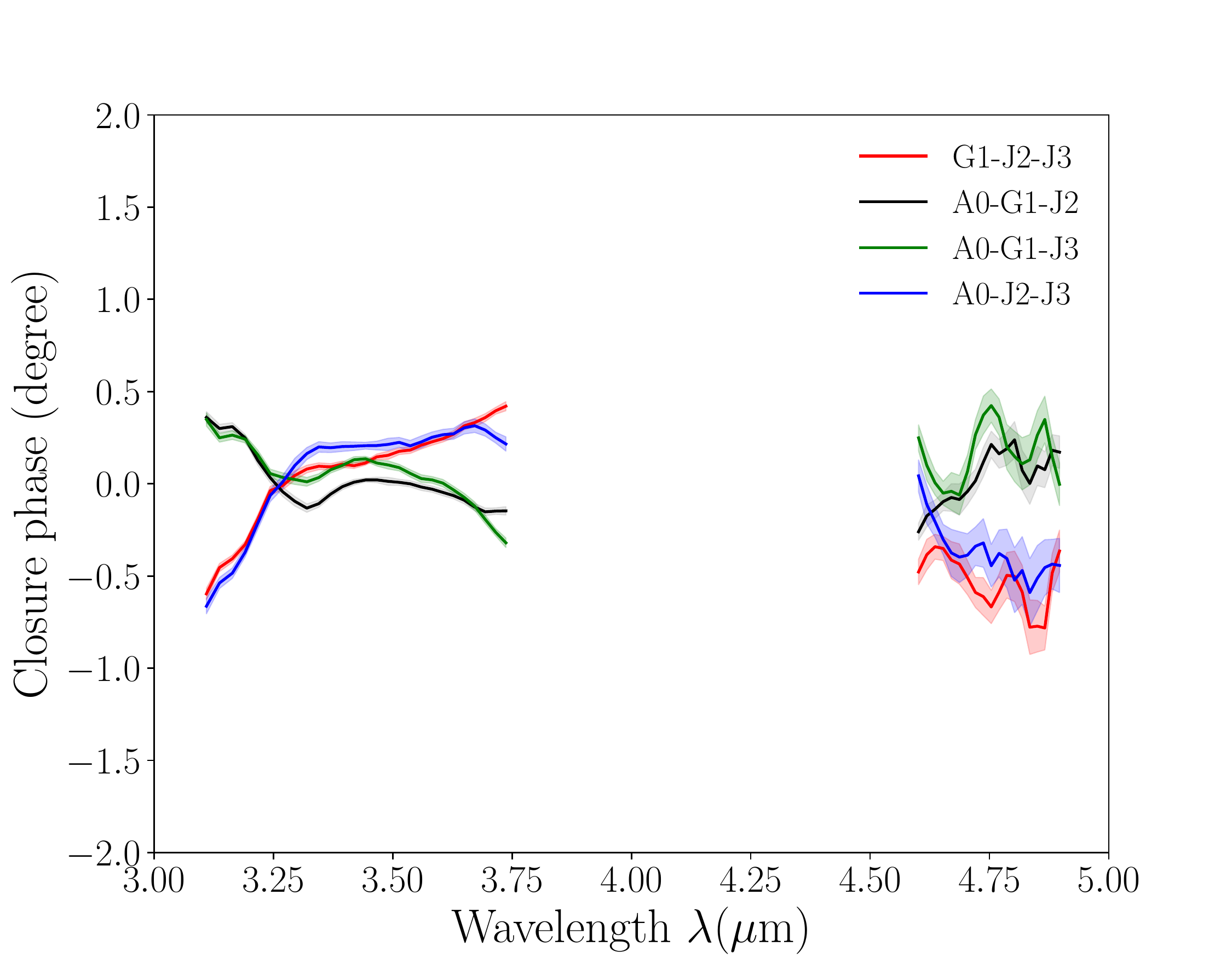}
  \caption{\#1 : $LM$ bands}
  \label{fig:1}
\end{subfigure}\hfil 
\begin{subfigure}[b]{.5\textwidth}
  \includegraphics[width=\linewidth]{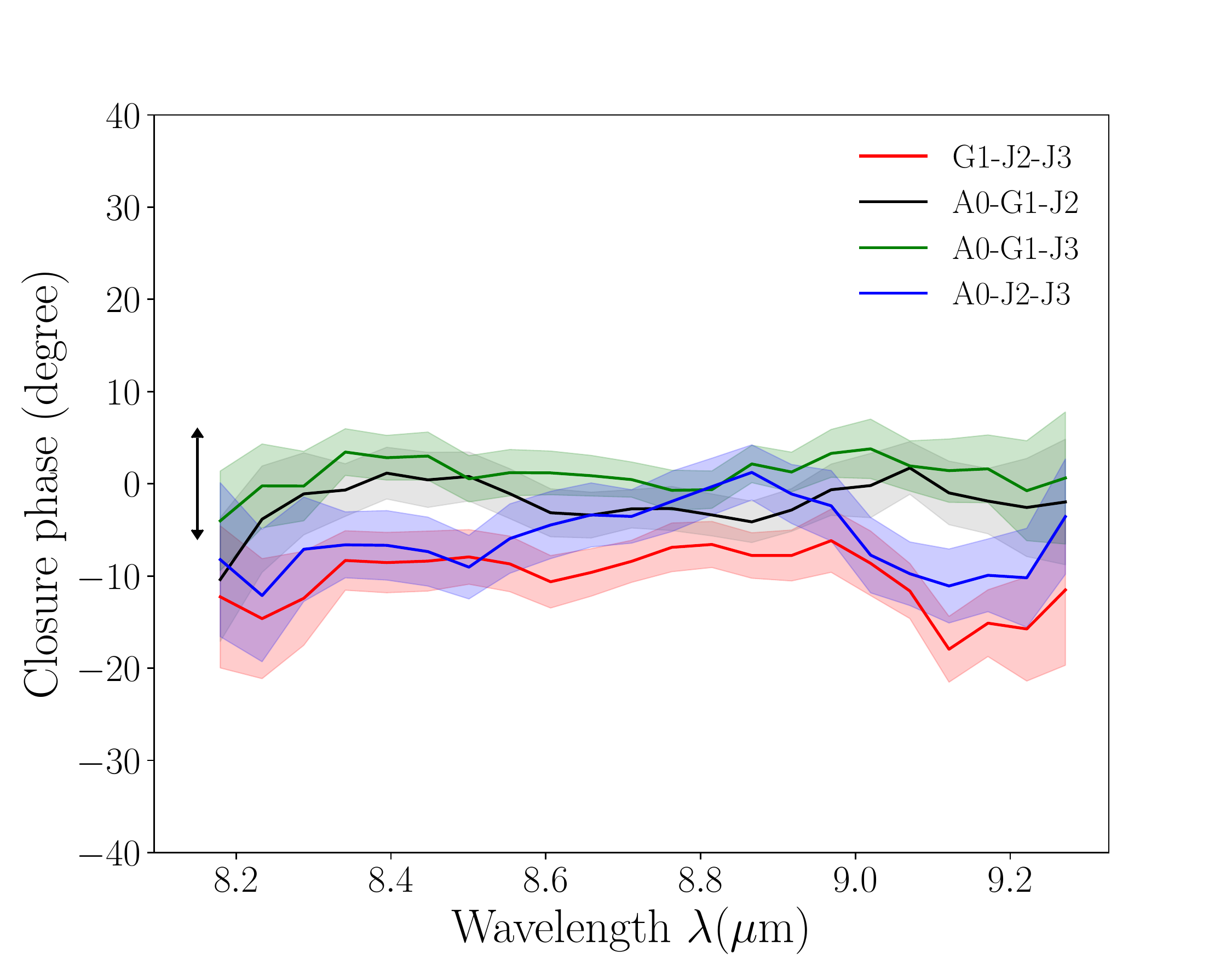}
  \caption{\#1 : $N$ band}
  \label{fig:4}
\end{subfigure}

\begin{subfigure}[b]{0.50\textwidth}
  \includegraphics[width=\linewidth]{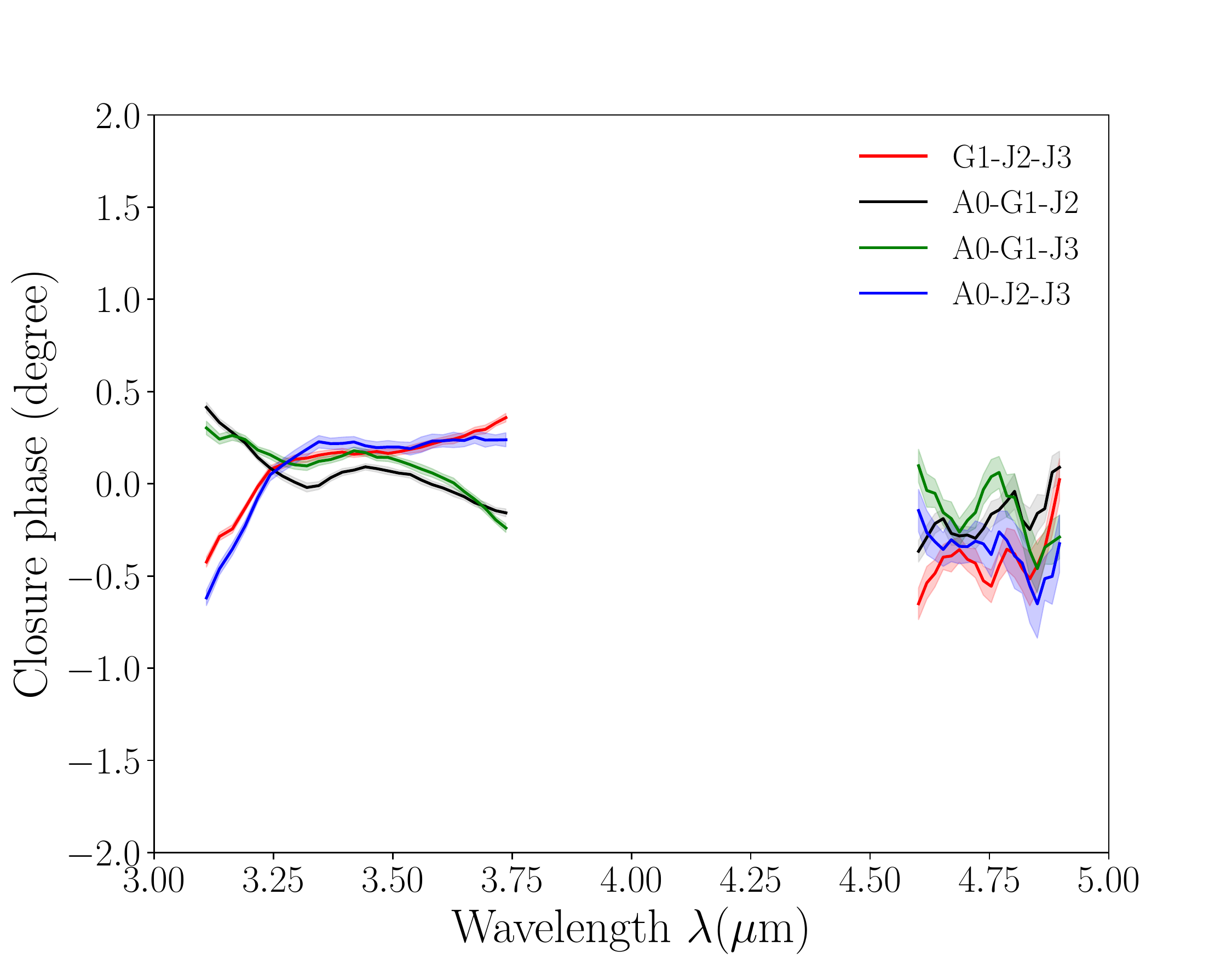}
  \caption{\#2 : $LM$ bands}
  \label{fig:2}
\end{subfigure}\hfil 
\begin{subfigure}[b]{.5\textwidth}
  \includegraphics[width=\linewidth]{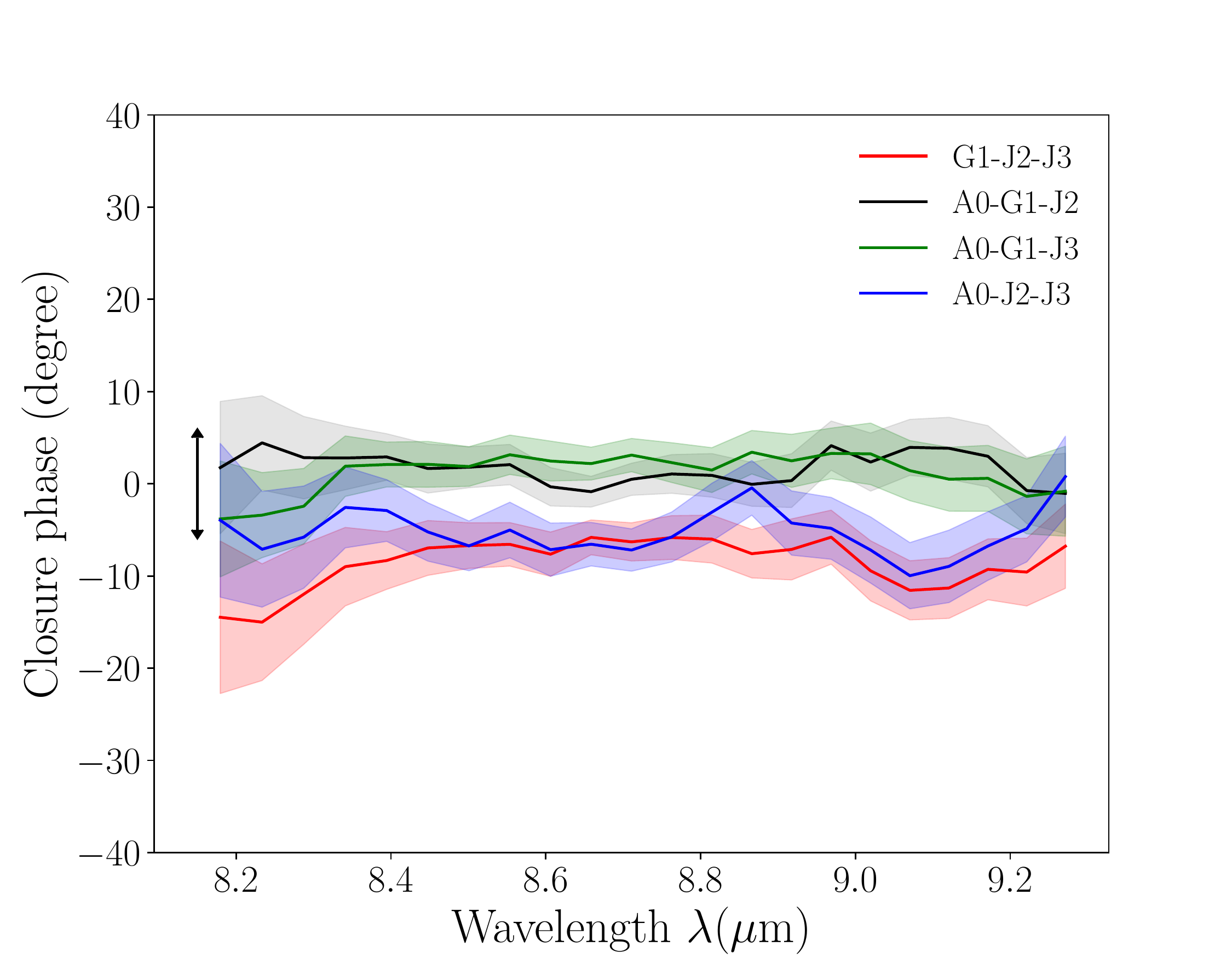}
  \caption{\#2 : $N$ band}
  \label{fig:5}
\end{subfigure}

\begin{subfigure}[b]{0.5\textwidth}
  \includegraphics[width=\linewidth]{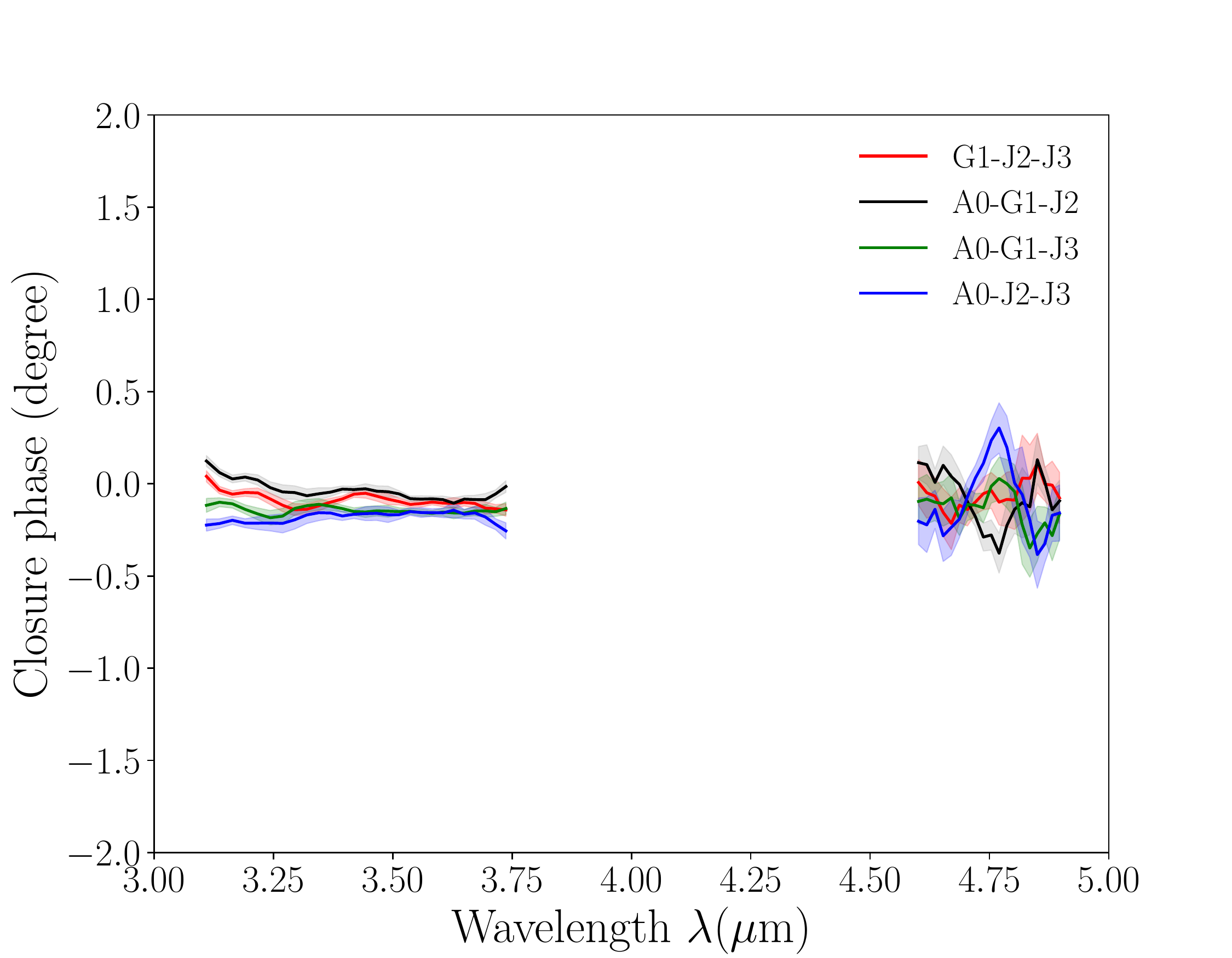}
  \caption{\#3 : $LM$ bands}
  \label{fig:3}
\end{subfigure}\hfil
\begin{subfigure}[b]{0.5\textwidth}
  \includegraphics[width=\linewidth]{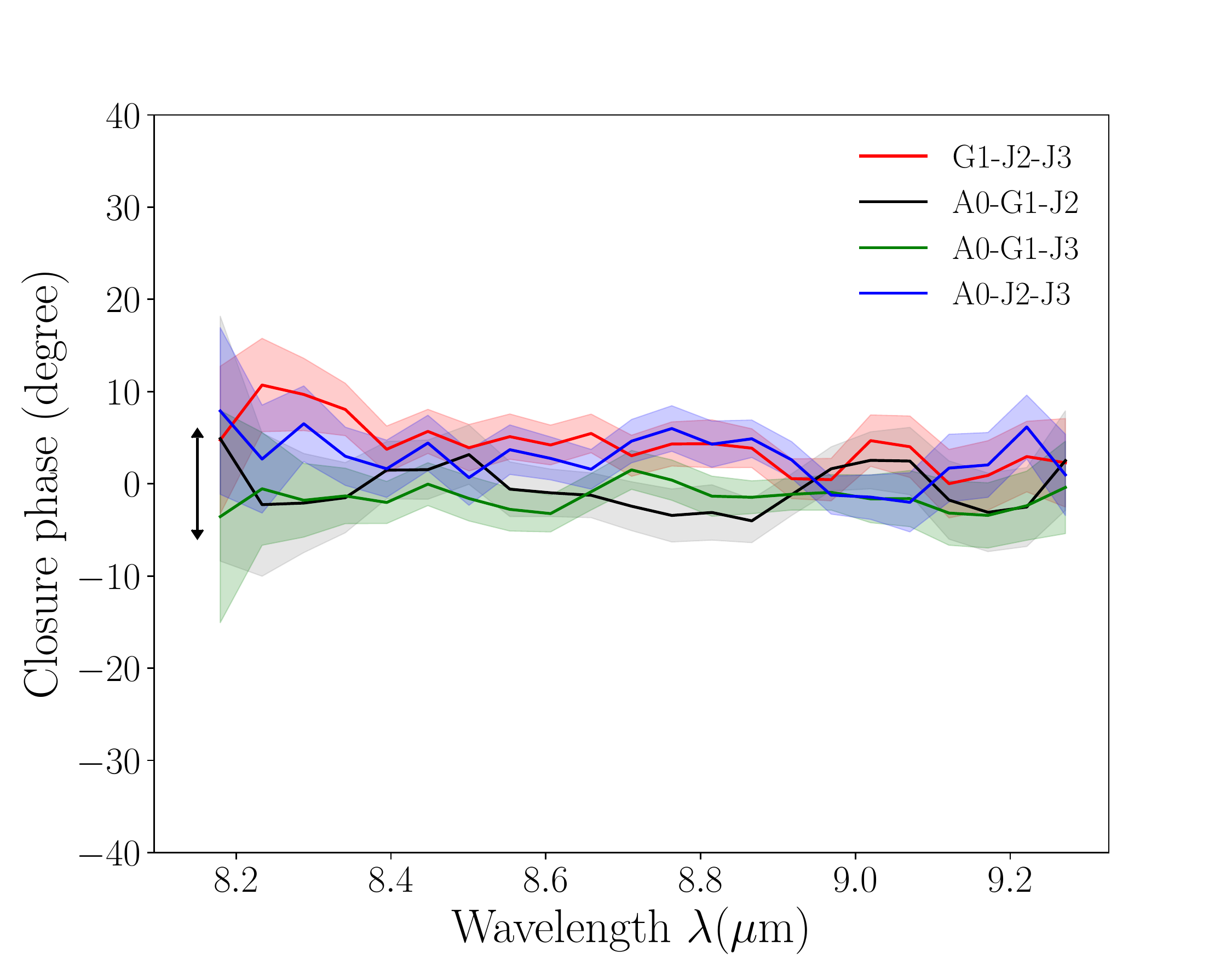}
  \caption{\#3 : $N$ band}
  \label{fig:6}
\end{subfigure}
\caption{Closure phase for observations \#1, 2 and 3 for $LM$ and the $N$ bands for each ATs triplet. The double arrows in the $N$ band panels represent the typical peak-to-valley variation.}
 \label{fig:t3}
\end{figure*}

\paragraph{}
MATISSE also provides closure phase measurements, which contain information about the spatial centro-symmetry of the brightness distribution of the source. For all the  closure phase measurements, we find an average closure phase of about 0$^\circ$ in the $L$, $M$ and the $N$ band from 8.2 to 9.25$\,\mu$m (see Fig.~\ref{fig:t3}). In the $N$ band we indicated the typical peak-to-valley dispersion of the MATISSE closure phase measurements with an error bar ($\approx$10$\,$degrees) and we discarded the spectral region beyond 10$\,\mu$m due to extremely large uncertainties (see Fig.~\ref{fig:t3}). Down to a sub-degree level (resp. 10 degrees level), our closure phase measurements are consistent with the absence of significant brightness spatial asymmetries in the environment around $\ell~$Car in $L$ and $M$ bands (resp. $N$ band). That justifies our use of centro-symmetric models in the following of the paper.

\begin{figure}[H]
\begin{center}
\includegraphics[width=0.53\textwidth]{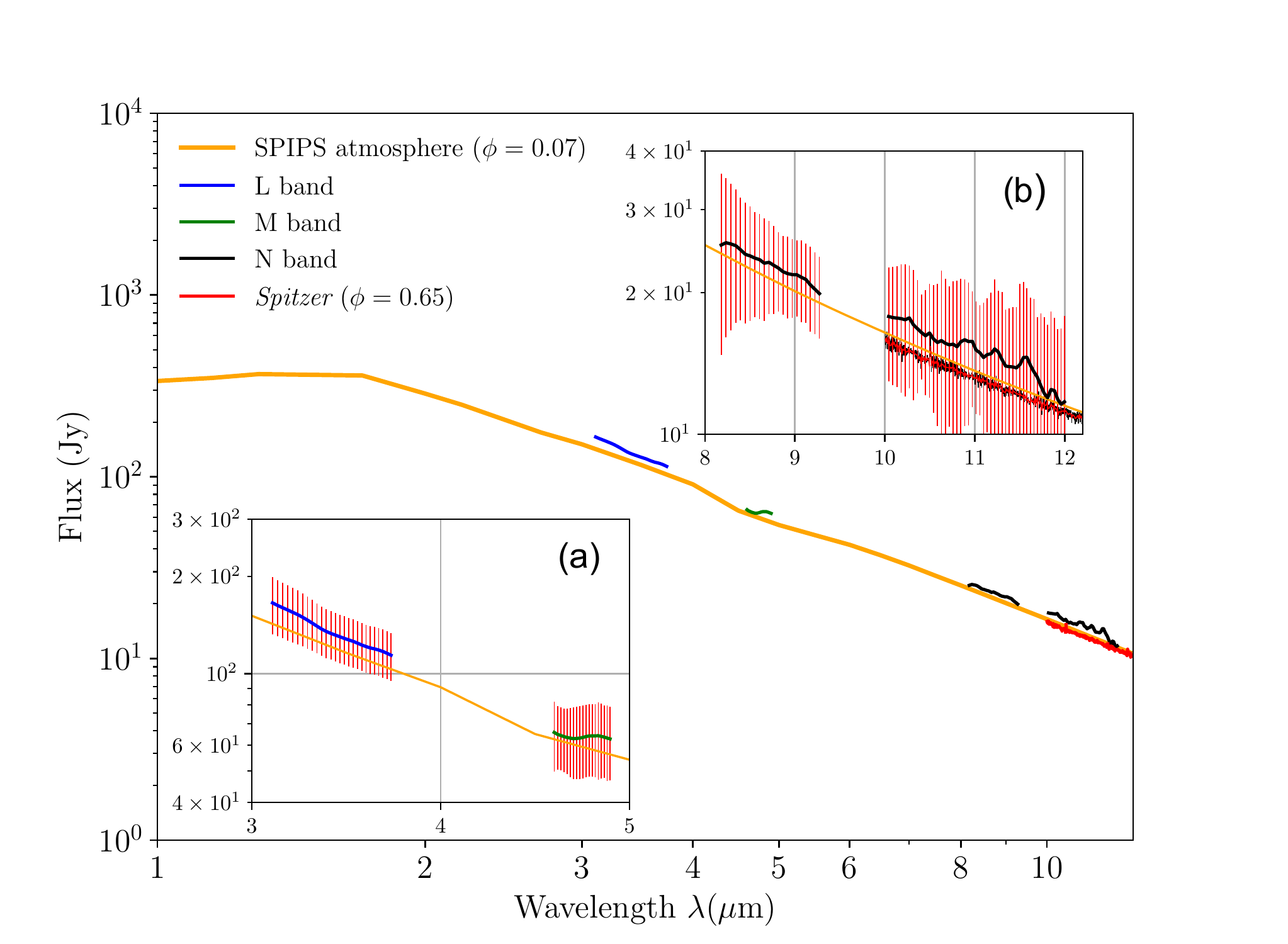}
\caption{\small Averaged calibrated total flux for observations $\#1$, 2 and 3 in $LM$ bands plus the correlated flux in $N$ band for observation $\#3$. Panels (a) and (b) refer to $LM$ and $N$ bands respectively. \texttt{SPIPS} photosphere model is interpolated at the phase corresponding to the MATISSE the observation with parameters: $T_\mathrm{eff}(\phi_\mathrm{obs})$=$5014\pm15\,$K;  log $g(\phi_\mathrm{obs})$=$0.96\pm0.10$;  and $\theta_\mathrm{UD}(\phi_\mathrm{obs})$=$2.887\pm0.003\,$mas. The \textit{Spitzer} spectrum used in Sect.~\ref{sect:spitzer} is plotted for comparison.\label{fig:total_flux}}

\end{center}
\end{figure}

\subsection{Flux calibration}\label{sect:flux_cal}

In $L$ and $M$ bands we calibrate the total flux $F_\mathrm{tot,sci}$ of $\ell$ Car using
\begin{equation}\label{eq:tot_flux_cal}
F_\mathrm{tot,sci}=\frac{I_\mathrm{tot,sci}}{I_\mathrm{tot,cal}} \times F_\mathrm{tot,cal}
\end{equation}
\noindent where $I_\mathrm{tot,sci}$ and  $I_\mathrm{tot,cal}$ are the observed total raw flux of the science target and the calibrator respectively. $F_\mathrm{tot,cal}$ is the known flux of the standard star. We calibrate $\ell$ Car using standard stars with the closest air mass which are q Car and $\beta$ Vol for the observations of the first and second night respectively. $F_\mathrm{tot,cal}$ is given by their atmospheric templates from \cite{Cohen1999}. Since the air mass of both target and calibrator are comparable we do not correct for the air mass. Moreover, while a chromatic correction exists for the $N$ band \citep{Sterzik2005} it is not calibrated (to our knowledge) for the $L$ and $M$ bands. Then we average the three observations, adding uncertainties and systematics between each measurement in quadrature.

Since $\ell$~Car is too faint in $N$ band for accurate photometry (and thus absolute visibility) measurements, we use the correlated flux for the $N$ band, which is the flux contribution from the spatially unresolved structures of the source. We calibrated the correlated flux of the science target $F_\mathrm{corr,sci}$ following :

\begin{equation}\label{eq:corr_flux_cal}
F_\mathrm{corr,sci}=\frac{I_\mathrm{corr,sci}}{I_\mathrm{corr,cal}} \times F_\mathrm{tot,cal} \times V_\mathrm{cal}
\end{equation}

\noindent where $I_\mathrm{corr,sci}$ and  $I_\mathrm{corr,cal}$ are the observed raw correlated flux of the science target and the calibrator respectively, and $V_\mathrm{cal}$ the calibrator visibility. For such an absolute calibration of the correlated flux, we need a robust interferometric calibrator with an atmospheric template given by \cite{Cohen1999}. Only $\beta$ Vol, observed during the second night, meets these two requirements. Thus we only calibrate the correlated flux measurements from observation $\#3$ using the known flux of $\beta$ Vol. The calibrated correlated fluxes from the different baselines do not present any resolved features, allowing us to average the correlated flux from the different baselines. 
Note that we have discarded the spectral region between 9.3 and 10$\,\mu$m, due to the presence of a telluric ozone absorption feature around 9.6$\,\mu$m that strongly impacts the data quality. The calibrated total fluxes in $L$ and $M$ bands and the averaged calibrated $N$ band correlated flux are presented in Fig.~\ref{fig:total_flux}. 

\paragraph{}
We draw here several intermediate conclusions.
First, the agreement between the measured flux with MATISSE, \textit{Spitzer} and the \texttt{SPIPS} atmosphere model of $\ell$ Car is qualitatively satisfactory within the uncertainties. However, the MATISSE flux measurements have rather large uncertainties ($\geq$ 10\%), making it difficult to determine the IR excess of $\ell$ Car with a precision at the few percent level. 
Secondly, we can see that the $N$ band correlated flux follows a pure Rayleigh-Jeans slope. This indicates the absence of silicate emission, which is consistent with the \textit{Spitzer} data (Sect.~\ref{sect:spitzer}). This is also in agreement with previous MIDI spectrum given by \cite{kervella09}.
Third, a resolved environment around $\ell$~Car is seen in the visibility measurements in the $L$ band. Indeed, the calibrated visibilities are significantly lower than the visibilities corresponding to a model of the star without CSE, as derived from the \texttt{SPIPS} algorithm at the specific phase of VLTI/MATISSE observations (see Fig.~\ref{fig:cal_sci}). Since the interferometric closure phase is zero the resolved structure around the star is centro-symmetric.
In the next section we apply a centro-symmetric model of envelope on the observed $V^2$ measurements in $L$ band in order to investigate the diameter and flux contribution of the envelope.

\section{Gaussian envelope model}\label{sect:envelope_model}
\begin{figure}[H]
\begin{center}
\includegraphics[width=0.53\textwidth]{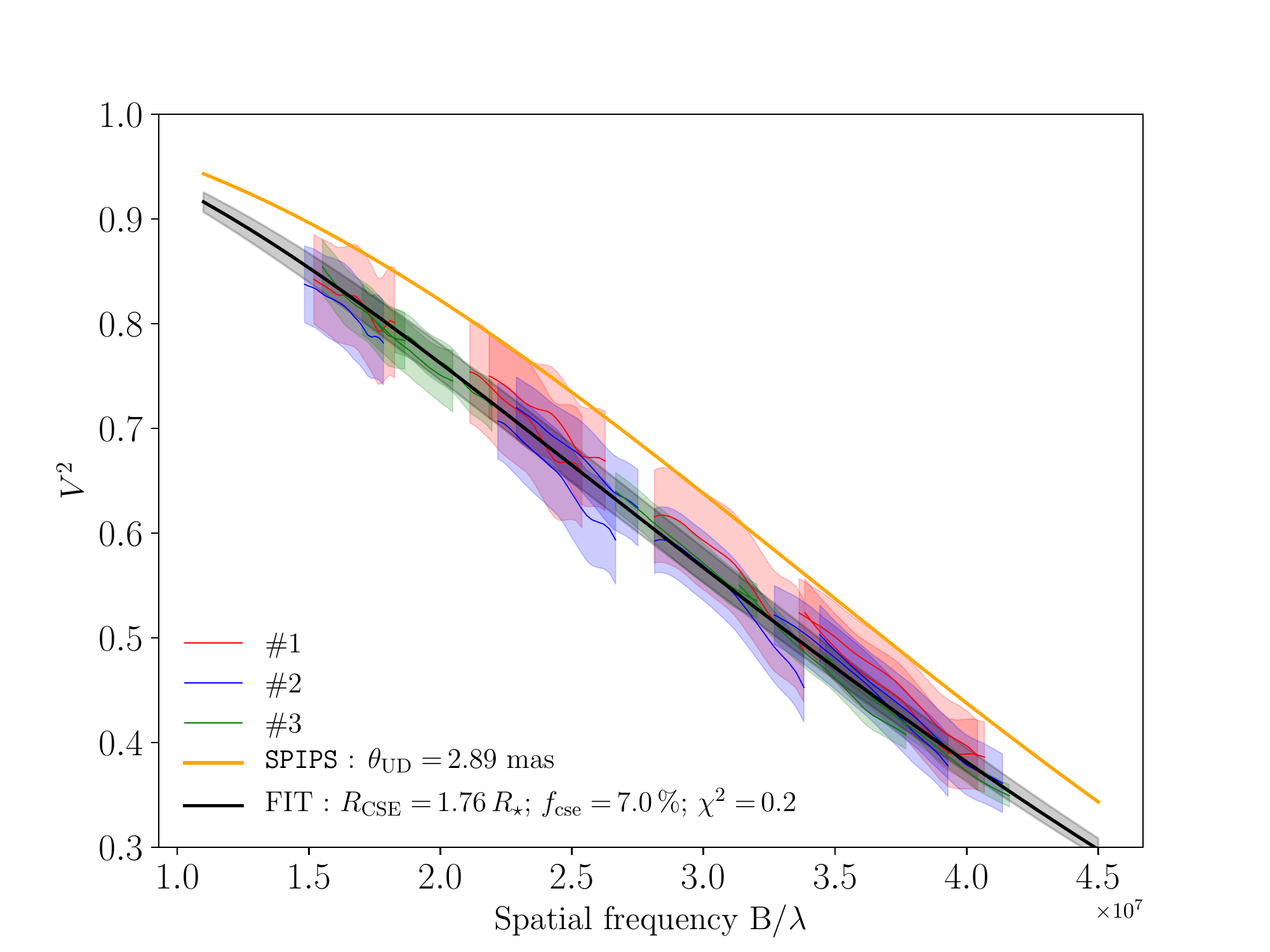}
\caption{\small Fitted Gaussian CSE around $\ell$ Car for the combined observations $\#1$, 2 and 3 in the $L$ band. The error on the visibility model is obtained using the covariance matrix of the fitting result.} 
        \label{fig:model_total}
\end{center}
\end{figure}

In this section we fit a geometrical model on the measured $L$ band visibilities. $\ell$ Car and its CSE are modeled with an uniform disk (UD) and a surimposed Gaussian distribution, as in the previous studies on $\ell$ Car CSE \citep{kervella06a,kervella09}. The UD diameter corresponds to the one at the specific phase of the observation derived by \texttt{SPIPS} that is $\theta_\mathrm{UD}=2.887\pm0.003\,$mas in the $L$ band. To model the CSE, we superimposed a Gaussian intensity distribution centered on the stellar disk. This CSE Gaussian model has two parameters which are the CSE diameter $\theta_\mathrm{CSE}$ taken as the Full Width at Half Maximum (FWHM), and the CSE flux contribution normalized to the total flux. We then performed a reduced $\chi_r^2$ fitting to adjust the total squared visibility of the model $V^2_\mathrm{tot}$ over the MATISSE observation in $L$ band. We computed the total squared visibility of the star plus CSE model using
\begin{equation}\label{eq:vis2}
    V_{\mathrm{tot}}^2(f)=\Big( F_\star \, \left \lvert V_\mathrm{UD}(f)\right \rvert + F_\mathrm{CSE} \,\left \lvert V_\mathrm{CSE}(f) \right \rvert \Big)^2
\end{equation}
 \noindent where $f$=$B_p$/$\lambda$ is the spatial frequency ($B_p$ the length of the projected baseline), $F_\star$ and  $F_\mathrm{CSE}$ are the normalized stellar and CSE flux contribution to the total flux respectively, normalized to unity $F_\star$ + $F_\mathrm{CSE}$ = 1. $V_\mathrm{UD}(f)$ is the visibility derived from the UD diameter of the star given by
\begin{equation}
    V_\mathrm{UD}(f)= 2\frac{J_1(\pi \theta_\mathrm{UD}f)}{\pi \theta_\mathrm{UD}f}
\end{equation}
where $J_1$ is the Bessel function of the first order. $V_\mathrm{CSE}(f)$ is the visibility of a Gaussian intensity distribution
\begin{equation}
    V_\mathrm{CSE}(f)=\exp \Big[ -\frac{(\pi\theta_\mathrm{CSE}f)^2}{4 \mathrm{ln}2}\Big].
\end{equation}
We perform the fit on observations $\#1$, 2 and 3 simultaneously, since $\ell$ Car is not expected to vary significantly between $\phi$=0.05 and 0.07. The derived visibility model is shown in Fig.~\ref{fig:model_total} and we present the best-fit parameters in Table~\ref{tab:fit}. The visibility model turns out to be consistent, within the error bars, with the three observations. However, the reduced $\chi_r^2$ is low (i.e. 0.2), thus the data are overfitted by the model and the uncertainties on the best-fit parameters are underestimated. In that case, to obtain more reliable uncertainties on the best-fit parameters we fit independently the three observations and we take the uncertainty as the standard deviation as of the resulted parameters.
 We resolve a CSE with a radius of 1.76$\pm0.28\,R_\star$ accounting for about 7.0$\pm1.4\%$ of the flux contribution in the $L$ band, that gives an IR excess of $-$0.07$\,$mag. 
Both the size and the flux contribution are in agreement with previous studies which have found a compact environment of size 1.9$\pm$1.4$\,R_\star$ in $K$ band with VINCI and 3.0$\pm$1.1$\,R_\star$ in $N$ band with MIDI \citep{kervella06a}, with few percents in flux contribution in both cases. As noted by \cite{kervella06a}, the large uncertainties in the $K$ band diameter is due to a lack of interferometric data for baselines between 15 to 75$\,$m. We also note that cycle-to-cycle amplitude variations discovered in $\ell$~Car \citep{anderson14,anderson16} could slightly affect the diameter in the $K$ band. On the other hand \cite{kervella09} found extended emission between 100 and 1000 AU ($\approx$100-1000$\,R_\star$) using MIDI and VISIR observations in the $N$ band. Thus it is possible that the CSE have both a compact and an extended component which are observable at different wavelengths.  The CSE flux contribution of 7\% is also in agreement with the MATISSE total flux within the uncertainties (see Fig.~\ref{fig:total_flux}), as well as the determined IR excess of about $-$0.05 to $-$0.1$\,$mag we derived from Fig.~\ref{fig:ir_excess} in Sect.~\ref{sect:spitzer}. 

\begin{table}
\begin{center}

\caption{\small Fitted parameters of a Gaussian CSE with flux contribution $F_\mathrm{CSE}(\%)$ and FWHM $\theta_\mathrm{CSE}$(mas) which is also expressed in stellar radius $R_\mathrm{CSE}$($R_\star$). Parameters $F_\mathrm{CSE}$ and $\theta_\mathrm{CSE}$ are weakly correlated with a correlation coefficient of 0.09. The uncertainties on the parameters are obtained using the standard deviation of the best-fit parameters for the three observations fitted independently. These results are compared with those obtained with VINCI in $K$ band and MIDI in $N$ band from \cite{kervella06a}.}\label{tab:fit}

\begin{tabular}{c|c|cc}
\hline
\hline
 &  This work & VINCI & MIDI \\
 \hline
 &  $L$ & $K$ & $N$\\
\hline  
$f_\mathrm{CSE}(\%)$      & 7.0$\pm$1.4 &4.2$\pm$0.2&-\\
\hline
$\theta_\mathrm{CSE}$(mas)      &  5.06$\pm$0.81 &5.8$\pm$4.5& 8.0$\pm$3.0   \\
$R_\mathrm{CSE}$($R_\star$)      &  1.76$\pm$0.28&1.9$\pm$1.4& 3.0$\pm$1.1  \\
\hline
$\chi_r^2$     &   0.2  & 0.65& -\\
\hline  
\end{tabular}
\normalsize
\end{center}
\end{table}

\section{Physical origin of the IR excess}\label{sect:discussion}
\subsection{Dust envelope}
The absence of emission features in the $N$ band in both MATISSE and \textit{Spitzer} spectra rules out the presence of dust from typical oxygen rich star mineralogy with amorphous silicates or aluminum oxide, which present a characteristic spectral shape in $N$ band. A pure iron envelope, which presents a continuum emission, is also unlikely to be created. We already considered then rejected these possibilities in Paper I (see the Figures 8 and 9 in it). We note that the presence of large grains of silicate, with a size of about 1$\,\mu$m and more, would lead to a broader emission which could completely disappear \citep{henning2010}. However the reason why large grains would preferentially be created in the envelope of Cepheids remains to be explained. In addition, considering the best-fit of the Gaussian CSE model, at a distance of 1.76$\,R_\star$ from the stellar center, i.e. 0.76$\,R_\star$ from the photosphere, the temperature would exceed 2000~K, therefore dust cannot survive since it would sublimate \citep{Gail1999}. These physical difficulties to reproduce the IR excess with dust envelopes were also recently pointed out by \cite{Gro2020} who fitted the SEDs of 477 Cepheids (including $\ell$ Car) with a dust radiative transfer code.

\subsection{A thin shell of ionized gas}
\begin{figure}[H]
\begin{center}
\includegraphics[width=0.53\textwidth]{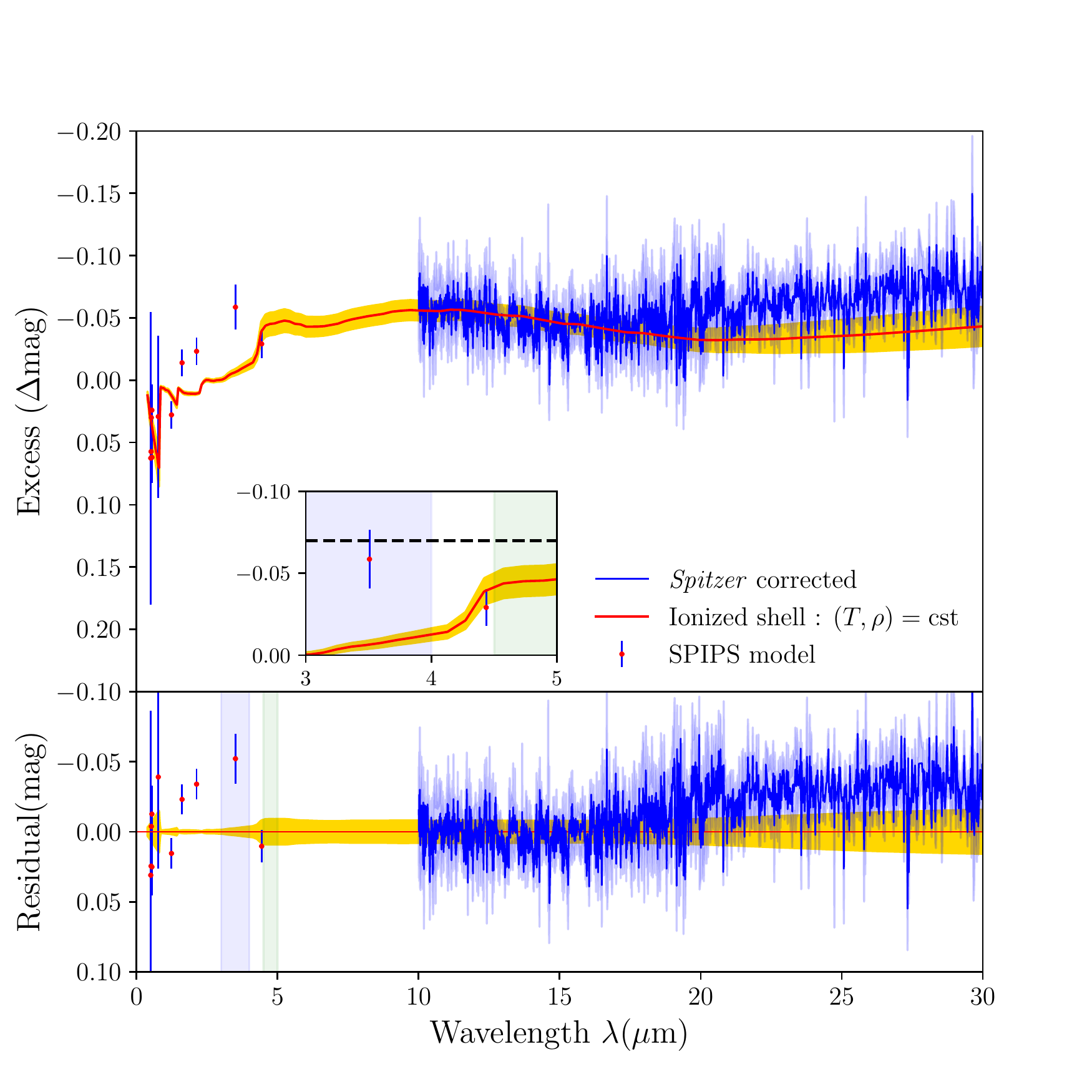}
\caption{\small $\ell$ Car IR excess fitting result of ionized gas shell following the method described in Paper I. Yellow region is the error on the magnitude obtained using the covariance matrix of the fitting result. We find a thin shell of ionized gas with radius $R_\mathrm{shell}$=1.13$\pm$0.02$\,R_\star $; temperature $T_\mathrm{shell}$=3791$\pm$85$\,K$; mass of ionized gas $M$=9.10$\times10^{-8}\pm$8.5$\times10^{-9}\,$\Msolar, using an ad-hoc offset of +0.03$\,$mag on the data in order to match the IR excess found by VLTI/MATISSE and the one of \texttt{SPIPS} (see the text); and  $\chi_r^2$=0.55. The subplot shows the comparison between the flux contribution derived from the Gaussian CSE (dashed line), the IR excess from \texttt{SPIPS} (red point) and IR excess from a shell of ionized gas in the $LM$ band (red line).}\label{fig:kromo}
\end{center}
\end{figure}

In Paper I we suggested the presence of a thin shell of ionized gas, with a radius of about 1.15$\,R_\star$, to explain the reconstructed IR excess of five Cepheids. We use the same model for $\ell$ Car with a shell of ionized gas, to test its consistency with IR excess and interferometric observations. In this part we perform a reduced $\chi_r^2$ fitting of the shell parameters on the IR excess reconstructed in Sect. \ref{sect:spitzer}. This simple model has three physical parameters which are: the shell radius; the temperature; and the mass of ionized gas. An additional parameter is applied as an offset to the whole reconstructed IR excess from Fig.~\ref{fig:ir_excess}, allowing the model to present either a deficit or an excess in the visible domain if necessary. Indeed, the key point is when considering a shell of ionized gas, this shell is expected to absorb the light coming from the star in the visible domain, which is currently not considered in the \texttt{SPIPS} algorithm when reconstructing the IR excess (the $V$ magnitude is forced to be zero, see Fig.~\ref{fig:excess-spips}). However, thanks to the VLTI/MATISSE measurements of flux contribution of the CSE in the $L$ band (7\%), we can fix this offset parameter consistently to 0.03 magnitude, to force the IR excess in $L$ band from \texttt{SPIPS} (that is $\approx$10\%) to be also 7\%, within the uncertainties. We present the result of the modeling ($\chi_r^2$=0.55) constrained by the spectrum of $\ell$ Car in Fig.~\ref{fig:kromo}. We find a thin shell radius of $R_\mathrm{shell}$=1.13$\pm0.02\,R_\star$ which is lower than the Gaussian CSE of 1.76$\pm0.28\,R_\star$ constrained by MATISSE observations. On the other hand this model also reproduces the \textit{Spitzer} spectrum better than the \texttt{SPIPS} predicted photometries. In particular its value in the $L$ band is about 1\% (see Fig.~\ref{fig:kromo}) which is not consistent with the value derived using $L$ band visibilities ($\sim 7$\%). This model has also derived weak near-infrared excess for several stars in Paper~I (see its Figure 11). This behaviour is physically explained by the free-bound absorption of hydrogen before the free-free emission dominates at longer wavelength.  We have explored the spatial parameters to test if a larger ionized envelope has the potential to match MATISSE observations with the IR excess. However, our model is not able to accurately reproduce all the observations. Also, even if our model is complex considering the physics involved (free-free and bound-free opacities description), its geometrical description is rather simple, with a constant density and temperature distribution. Thus, it might not be adapted to the modeling of a larger envelope.

\subsection{Limit of the model}
 The derived parameters of the shell of ionized gas depend on the initial IR excess derived by \texttt{SPIPS}. Any systematics in this calculation could in turn affect the IR excess and the parameters of the shell of ionized gas. As noted by \cite{Hocde2020a}, the derived distance of \texttt{SPIPS} would be systematically affected if the shell of ionized gas is not taken into account, because of its absorption in the visible. This systematic should be only few percents ($10^{\Delta \mathrm{m}/2.5}$) all parameters being unchanged. The direct implementation of the gas model into the \texttt{SPIPS} fitting has to be tested in forthcoming studies to correct this uncertainty. Moreover, Gallenne et al. 2021 (submitted) studied the ad-hoc IR excess derived by SPIPS for a larger sample of stars and found that angular diameter and the colour excess are the most impacted parameters when no ad-hoc IR excess model is included. However in the case of $\ell$~Car these parameters appear to be well constrained by angular diameter measurements, the reddening and the distance are also in agreement with values found in the literature (see Appendix \ref{app:spips}).

\subsection{Perspectives}
An interesting physical alternative is to consider free-free emission produced by negative hydrogen ion H$^-$, with free electrons provided by metals. Indeed for an envelope of 2$\,R_\star$ the average temperature is about 3000~K and the hydrogen is neutral. Thus, in that case, most of the free electrons would be provided by silicon, iron and magnesium which have a mean first ionization potential of 7.89~eV. As neutral hydrogen is able to form negative hydrogen ion, H$^-$ could generate an infrared excess with free-free emission. This phenomenon is known to produce a significant IR excess compared to dust emission in the extended chromosphere of cool supergiant stars such as Betelgeuse \citep{Gilman1974,Humphreys1974,lambert1975,Altenhoff1979,skinner1987}. Moreover, the photo-detachment potential occurs at 1.6$\,\mu$m for the H$^{-}$ mechanism, thus free-free emission should dominate the IR excess above this wavelength, as it is suggested by the \texttt{SPIPS} cycle-averaged IR excess. This single photo-detachment has also the advantage to avoid the important bound-free absorption in the near-infrared produced in the preceding model. We thus suggest that the compact structure we resolved around $\ell$ Car has the potential to produce IR excess through H$^-$ free-free mechanism by analogy with extended chromosphere of cool supergiant stars. Further investigations are necessary to confirm this hypothesis.

\section{Conclusions}\label{sect:conclusion}
Determining the nature and the occurrence of the CSEs of Cepheids is of high interest to quantify their impact on the Period-Luminosity relation, and also for understanding the mass-loss mechanisms at play. 
In this paper, we constrain both the IR excess and the geometry of the CSE of $\ell$~Car using both \textit{Spitzer} low-resolution spectroscopy and MATISSE interferometric observations in the mid-infrared. Assuming an IR excess \textit{ad-hoc} model, we used \texttt{SPIPS} to derive the photospheric parameters of the star in order to compare with \textit{Spitzer} and MATISSE at their respective phase of observation. This analysis allows to derive the IR excess from \textit{Spitzer} and also to derive the CSE properties in the $L$ band spatially resolved with MATISSE. These observations lead to the following conclusions on the physical nature of $\ell$ Car's CSE:
\begin{enumerate}
\item We resolve a centro-symmetric and compact structure in the $L$ band with VLTI/MATISSE that has a radius of about 1.76$\,R_\star$. The flux contribution is about 7\%.
\item We find no clear evidence for dust emission features around 10$\,\mu$m ($N$ band) in MATISSE and \textit{Spitzer} spectra which suggests an absence of circumstellar dust.
\item Our dedicated model of shell of ionized gas better reproduces the mid-infrared \textit{Spitzer} spectrum and implies a size for the CSE lower than the one derived from VLTI/MATISSE observations (1.13$\pm 0.02\,R_\star$ versus $1.76 \pm$0.28$\,R_\star$), as well as a lower flux in the $L$ band (1\% versus 7\%).
\item We suggest that improving our model of shell of ionized gas by including the free-free emission from negative hydrogen ion would probably help reproducing the observations, i.e. the size of the CSE and the IR excess, in particular in the $L$ band.
\end{enumerate}

While the compact CSE of $\ell$ Car is likely gaseous, the exact physical origin of the IR excess remains uncertain. Further observations of Cepheids depending on both their pulsation period and their location in the HR diagram are necessary to understand the CSE's IR excess. This could be a key to unbias Period-Luminosity relation from Cepheids IR excess.

\begin{acknowledgements}
The authors acknowledge the support of the French Agence Nationale de la Recherche (ANR), under grant ANR-15-CE31-0012- 01 (project UnlockCepheids). The research leading to these results has received funding from the European Research Council (ERC) under the European Union’s Horizon 2020 research and innovation program under grant agreement No 695099 (project CepBin). This research was supported by the LP2018-7/2019 grant of the Hungarian Academy of Sciences. This research has been supported by the Hungarian NKFIH grant  K132406.  This research made use of the SIMBAD and VIZIER\footnote{Available at \url{http://cdsweb.u- strasbg.fr/}} databases at CDS, Strasbourg (France) and the electronic bibliography maintained by the NASA/ADS system. 
This research also made use of Astropy, a  community-developed corePython package for Astronomy \citep{astropy2018}. Based on observations made with ESO telescopes at Paranal observatory under program IDs: 0104.D-0554(A).  This research has benefited from the help of SUV, the VLTIuser support  service of the Jean-Marie Mariotti Center (\url{http://www.jmmc.fr/suv.htm}). This research has also made use of the Jean-Marie Mariotti Center \texttt{Aspro}
service \footnote{Available at \url{http://www.jmmc.fr/aspro}}.
\end{acknowledgements}
\bibliographystyle{aa}  
\bibliography{bibtex_vh} 

\begin{appendix} 
\section{SPIPS data set and fitted pulsational model of the star sample.\label{app:spips}}
\begin{figure*}[h]
\begin{center}
\includegraphics[width=\hsize]{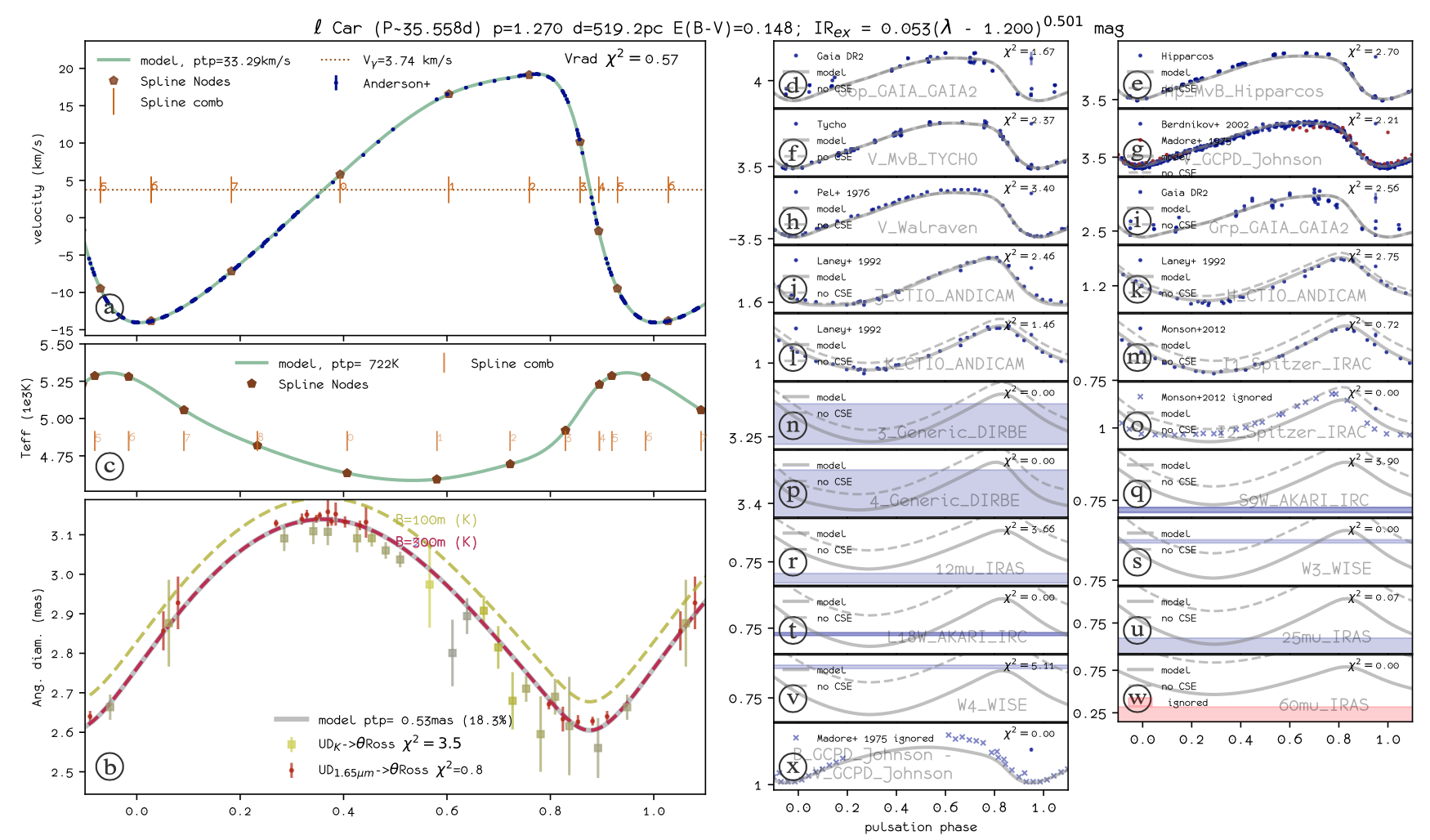}
\caption{\label{l_car_spips}\small The SPIPS results of $\ell$ Car. \textbf{Velocity:} \cite{anderson14}. \textbf{Effective temperature:} No data. \textbf{Angular diameter:} \cite{kervella06a}. \textbf{Photometry:} \cite{Madore1975}, \cite{Pel1976}, 
\cite{IRAS1984}, 
\cite{laney1992}, 
\cite{hipparcos1997},
\cite{berdnikov2002},
\cite{DIRBE2004},
\cite{tycho2006},
\cite{WISE2010},
\cite{AKARI2010},
\cite{Monson2012},
\cite{GaiaDR2}.}
\end{center}
\end{figure*}

Figure \ref{l_car_spips} is organized as follows:  pulsational velocity,  effective temperature and angular diameter curves according to the pulsation phase are shown on the left panels, while  the right panels display photometric data in various bands. Above the figure, the projection factor set to $p=1.270$ is indicated, along with the fitted distance $d$ using parallax-of-pulsation method, the fitted color excess E($B-V$), and the \textit{ad-hoc} IR excess law. In this work we arbitrarily set the p-factor to 1.27 following the general agreement around this value for several stars in the literature \citep{nardetto04,merand05,nardetto09}. However, there is no agreement for the optimum value to use, and it is also the case for $\ell$ Car. In the case of this star \citep{nardetto07,nardetto09} found a p-factor of 1.28$\pm$0.02 and 1.22$\pm$0.04 using different observational techniques which both are in agreement with 1.27 within about 1 sigma. Thus, a possible systematic cannot be excluded although the distance derived by SPIPS ($d=519.2\pm4.5\,$pc) is in agreement for example with  the distance obtained using \textit{Gaia} parallaxes (505$\pm$28$\,$pc) \citep{GaiaDR2} or the one derived using a recent period-luminosity calibrated for Cepheids in the Milky Way (544$\pm$32$\,$pc) \citep{Breuval2020}.

Gallenne et al. (2021, submitted) have shown that the absence of IR excess model in the \texttt{SPIPS} fit can affect the derivation of the angular diameter and the colour excess. In our fit using IR excess model, we emphasize the good agreement of the angular diameter model with the data (panel b), and also the consistency of the derived colour excess with the one given by \textit{Stilism} 3D extinction map \citep{2017stilism}, that is $0.148\pm0.006$ versus $0.116\pm0.26$. In the photometric panels, the gray dashed line corresponds to the magnitude of the SPIPS model without CSE. It actually corresponds to the magnitude of a Kurucz atmosphere model, $m_\mathrm{kurucz}$, obtained with the ATLAS9 simulation code from \cite{castelli2003}, with solar metallicity and a standard turbulent velocity of 2~km/s. These models are indeed suitable for deriving precise synthetic photometry \citep{Castelli1999}, in comparison with MARCS models giving rough flux estimates\footnote{\url{https://marcs.astro.uu.se/}}. The gray line corresponds to the best SPIPS model, which is composed of the latter model without CSE plus an IR excess model. In the angular diameter panels, the gray curve corresponds to limb-darkened (LD) angular diameters.
For stars with solar metallicity, when effective temperature is low enough, CO molecules can form in the photosphere and absorb light in the CO band-head at 4.6 $\mu m$ \citep{scowcroft2016}. This effect is observed in the \textit{Spitzer} I2 IRAC dataset. In this case, these data were ignored during the fitting of SPIPS. Panels presenting a horizontal blue bar contain only one data with undetermined phase of observation, thus these photometric mid-infrared bands were not used in the IR excess reconstruction in Sect.~\ref{sect:spips_photometry} and \ref{sect:spitzer}

\section{Verification of the diameters of the calibrators}\label{sect:b_cen}

\begin{figure*}[t!] 

\begin{subfigure}{0.48\textwidth}
\includegraphics[width=\linewidth]{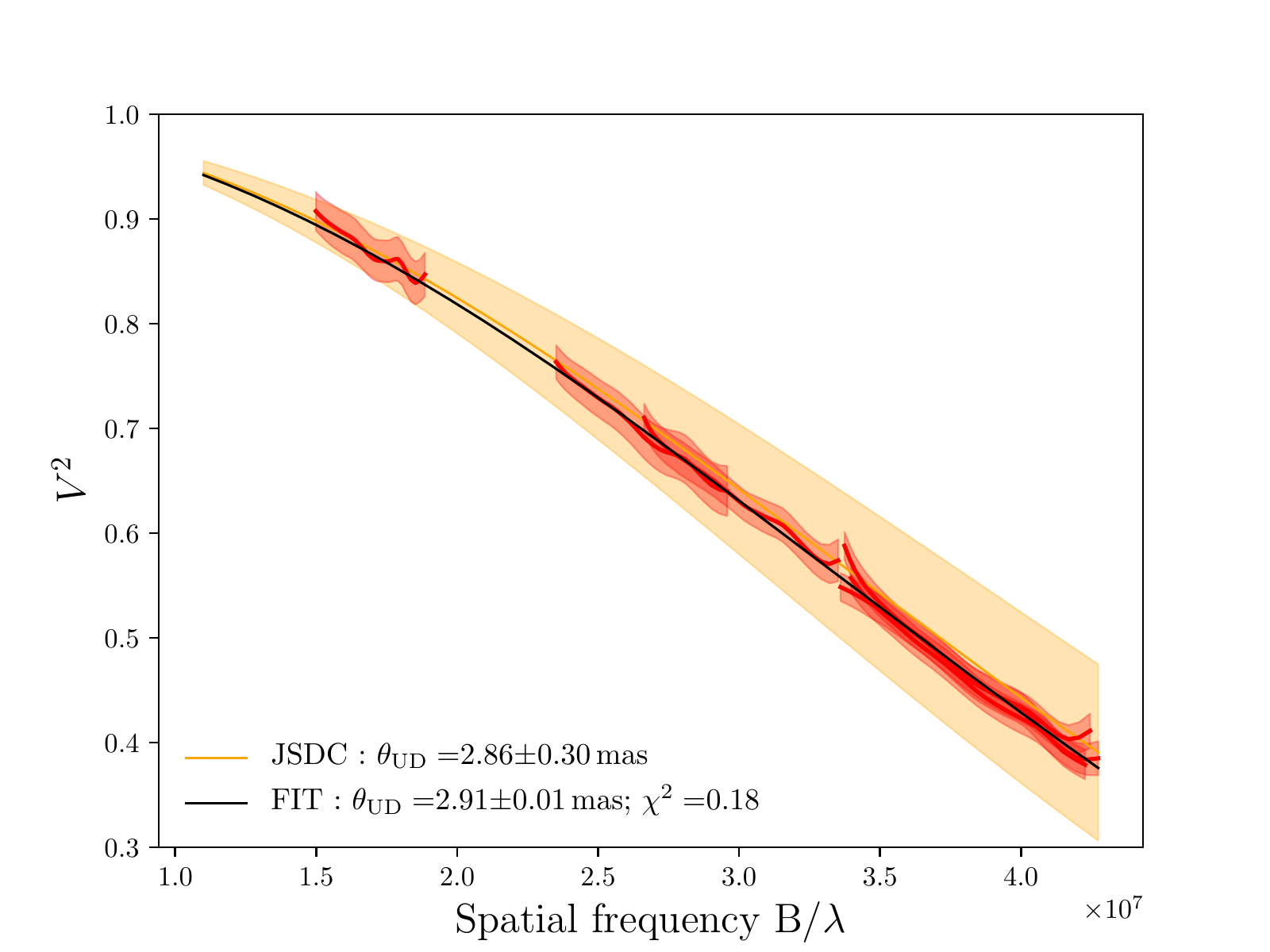}
\caption{$\epsilon$ Ant calibrated with e Cen} \label{fig:eps_ant_e_cen}
\end{subfigure}\hspace*{\fill}
\begin{subfigure}{0.48\textwidth}
\includegraphics[width=\linewidth]{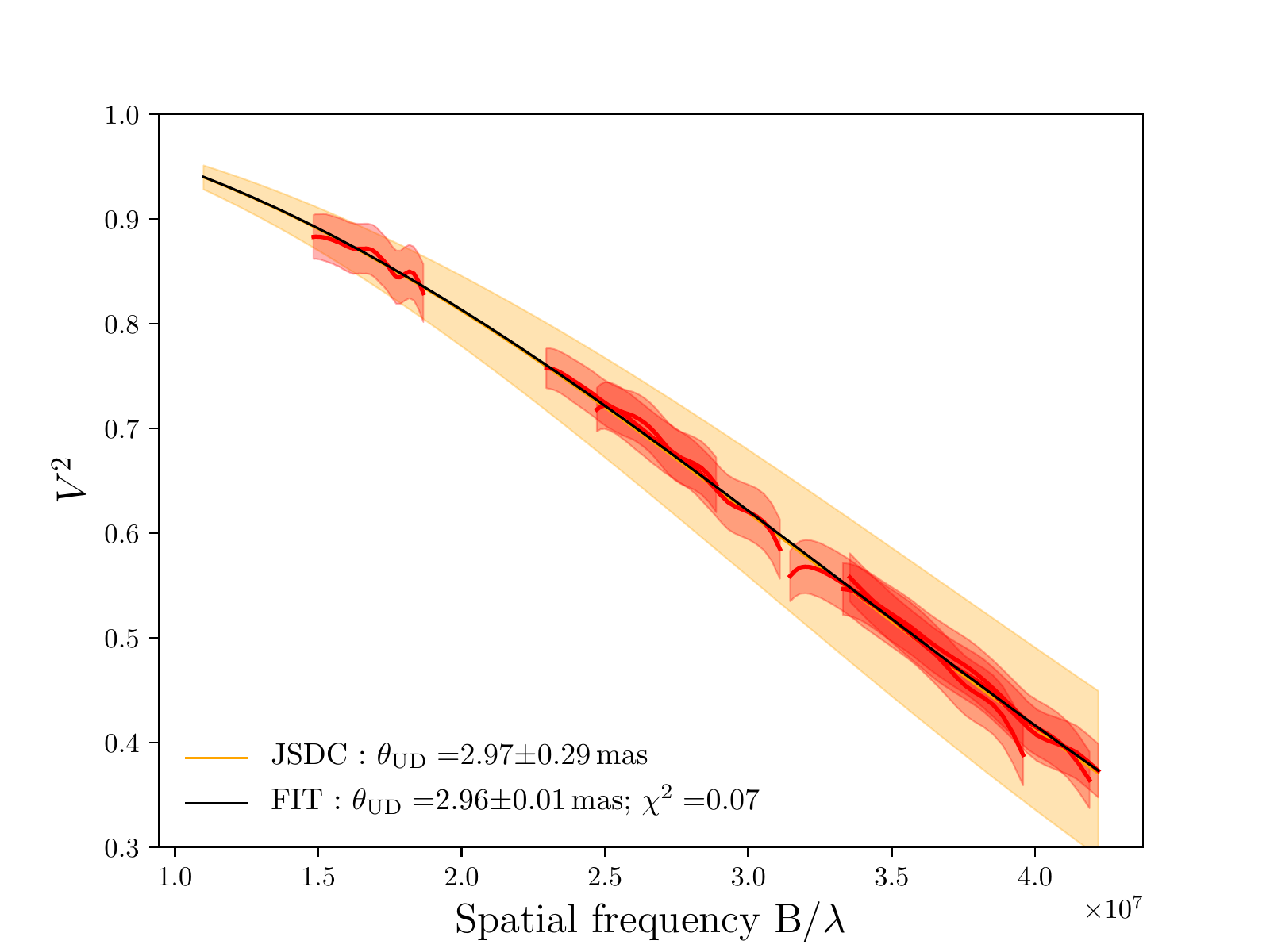}
\caption{e Cen calibrated with $\epsilon$ Ant} \label{fig:e_cen_eps_ant}
\end{subfigure}

\medskip
\begin{subfigure}{0.48\textwidth}
\includegraphics[width=\linewidth]{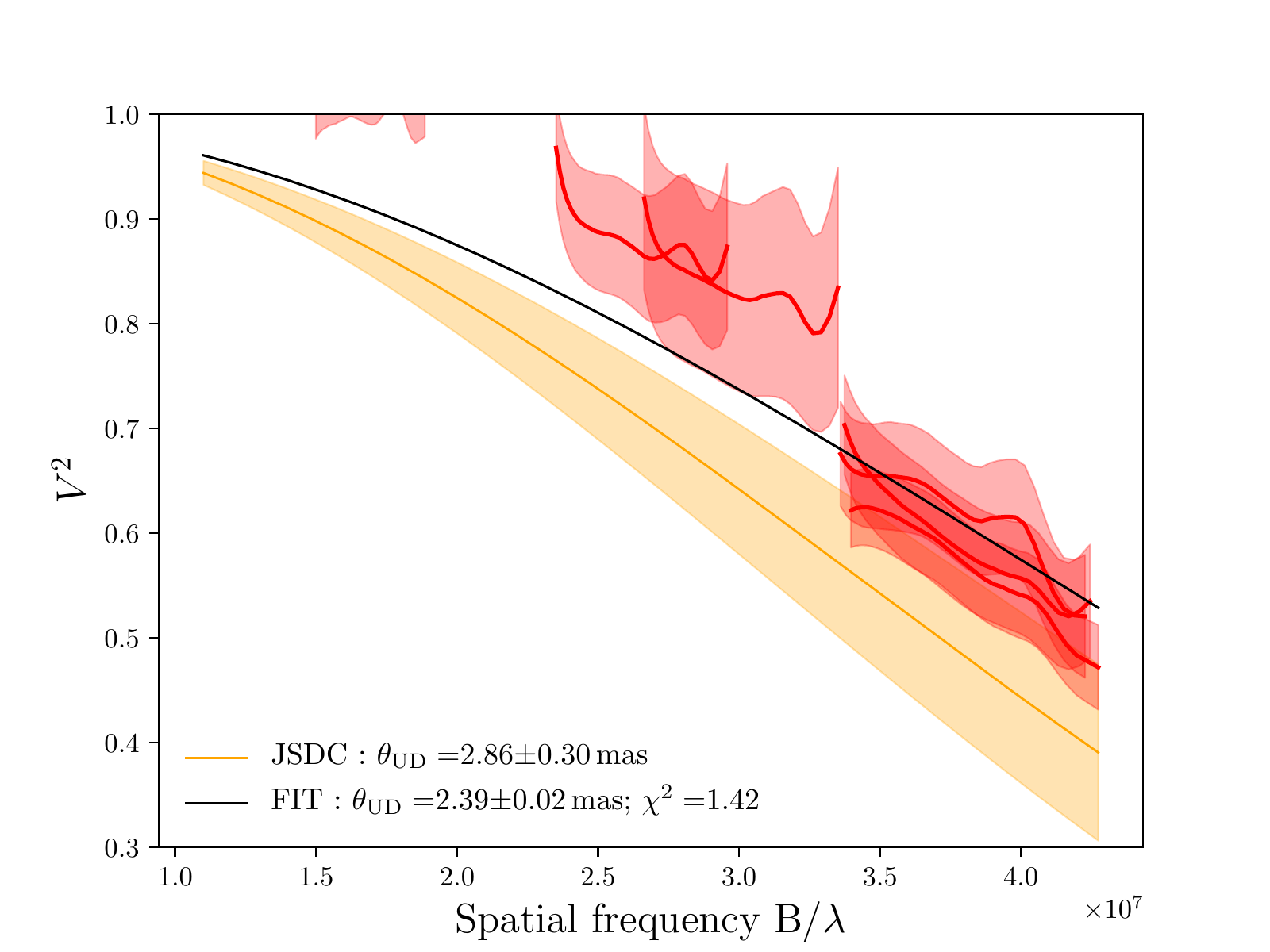}
\caption{$\epsilon$ Ant calibrated with B Cen} \label{fig:e_eps_ant_b_cen}
\end{subfigure}\hspace*{\fill}
\begin{subfigure}{0.48\textwidth}
\includegraphics[width=\linewidth]{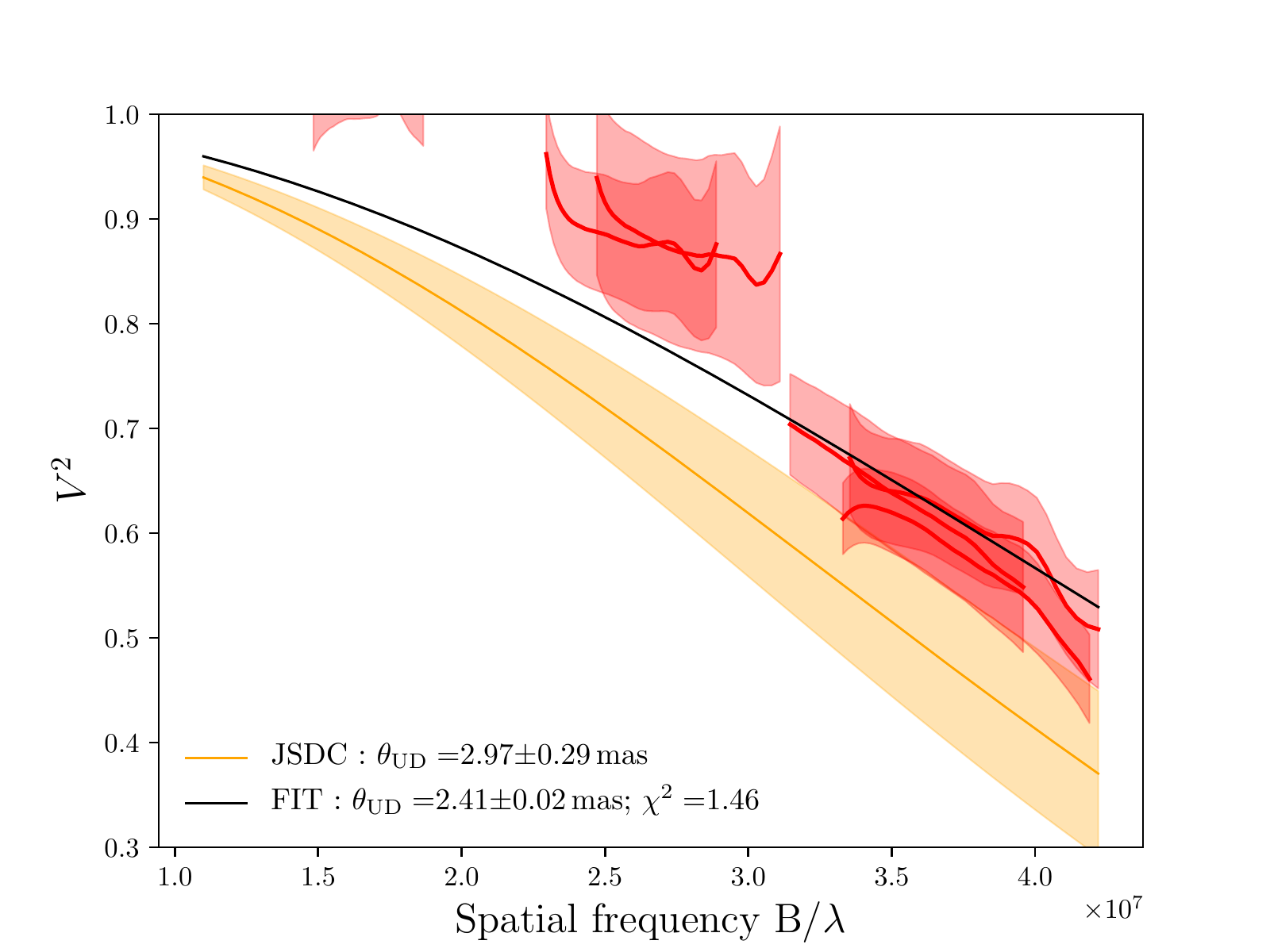}
\caption{e Cen calibrated with B Cen} \label{fig:e_cen_b_cen}
\end{subfigure}

\medskip
\begin{subfigure}{0.48\textwidth}
\includegraphics[width=\linewidth]{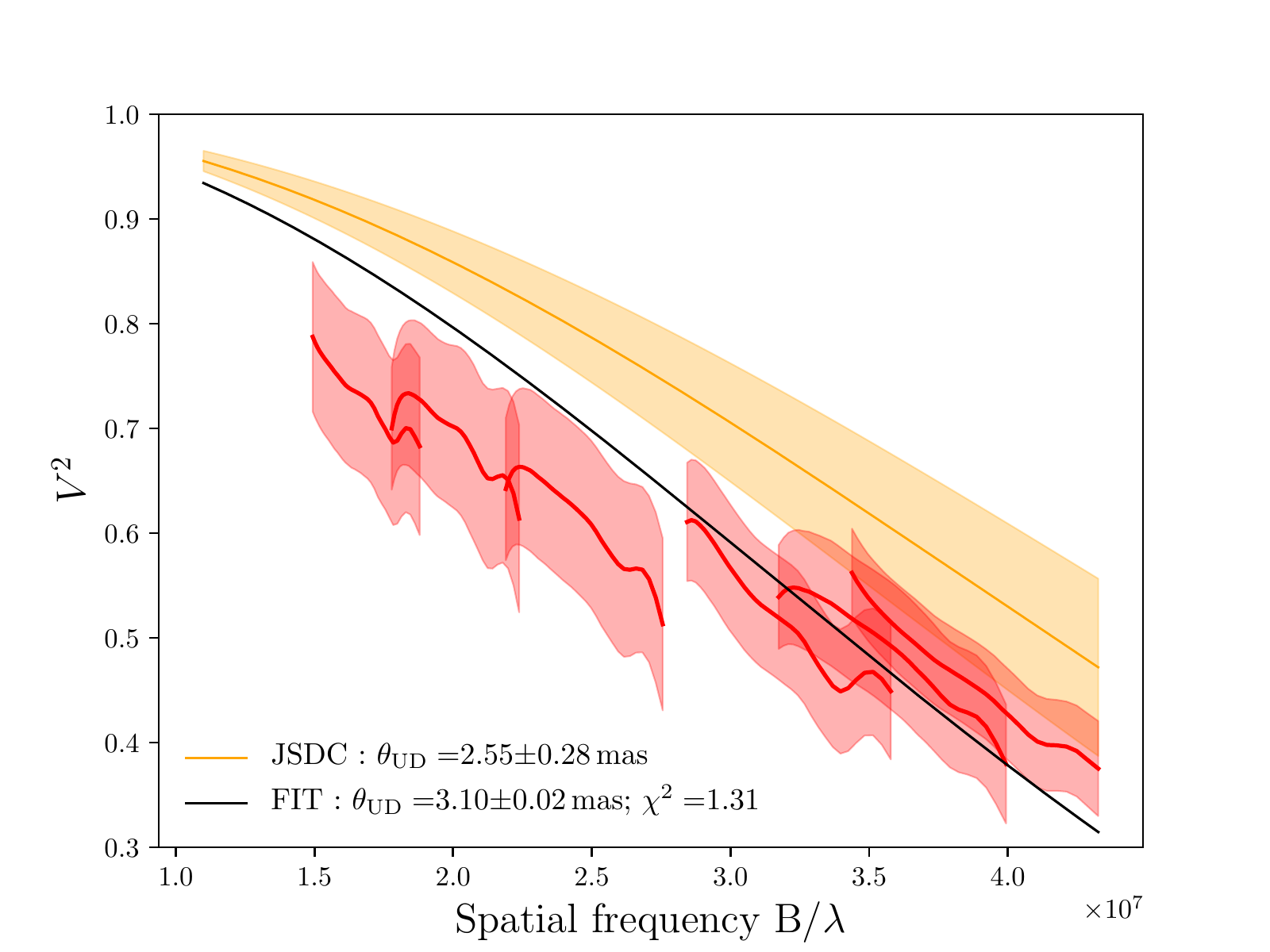}
\caption{B Cen calibrated with $\epsilon$ Ant} \label{fig:b_cen_eps_ant}
\end{subfigure}\hspace*{\fill}
\begin{subfigure}{0.48\textwidth}
\includegraphics[width=\linewidth]{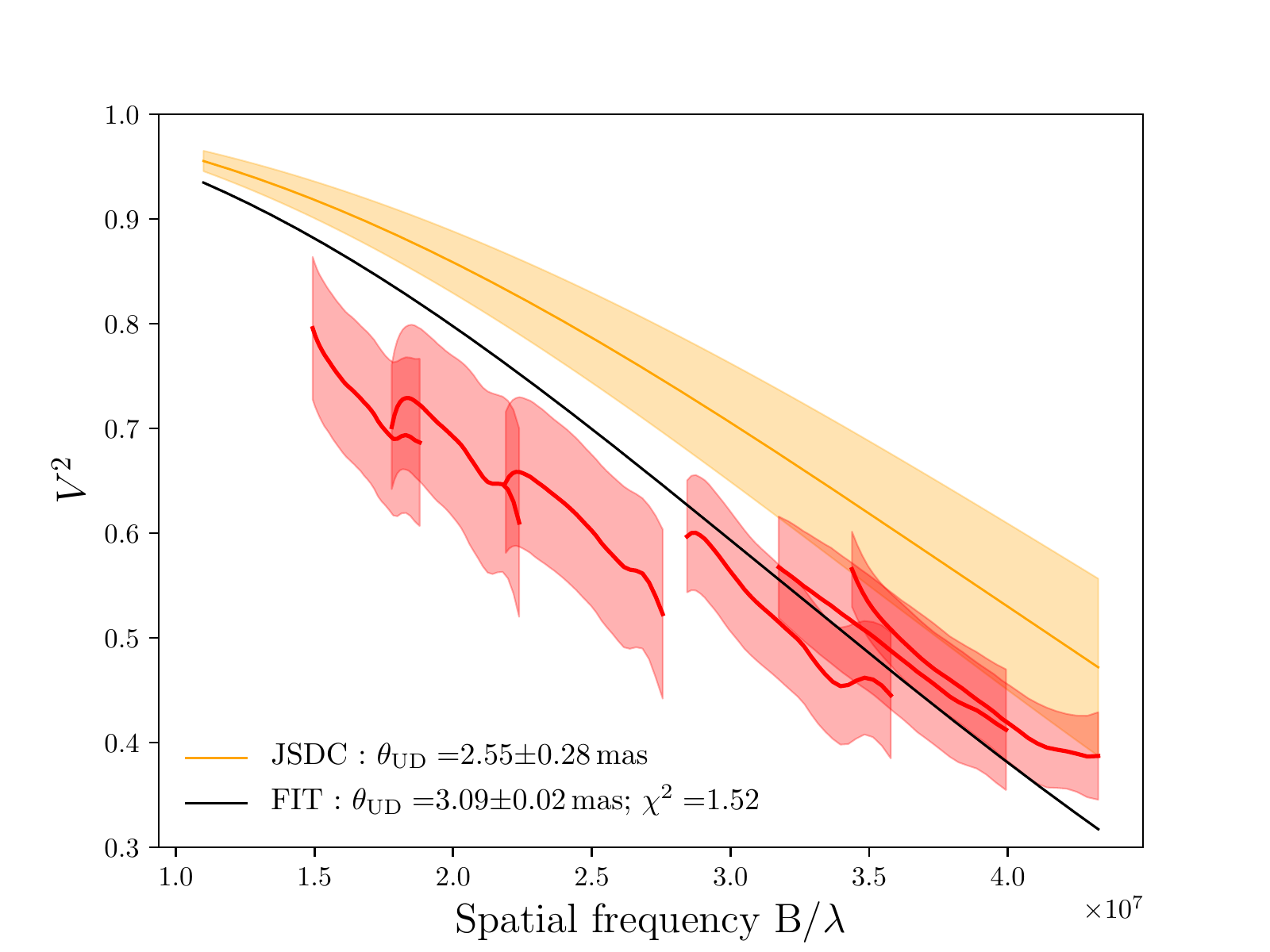}
\caption{B Cen calibrated with e Cen} \label{fig:b_cen_e_cen}
\end{subfigure}

\caption{\small Inter-calibration of the different calibrators and comparison with JSDC diameters. (a) and (b) : $\epsilon$ Ant and e Cen diameters are in agreement with the JSDC when they are calibrated from each other. (c) and (d) : On the contrary the diameter of $\epsilon$ Ant and e Cen are inconsistent with the JSDC catalogue when they are calibrated with B Cen. (e) and (f) : B Cen diameter is inconsistent with the JSDC catalogue.}\label{fig:inter_calibs}
\end{figure*}

In order to verify the consistency of the standard stars of the first night (27 February), we consider all calibrators available during the night (B~Cen, e Cen, $\epsilon$ Ant) and we inter-calibrate them. Thus, each calibrator is calibrated using the two others. We performed six different calibrations in total, following Table~\ref{Tab.inter_cal} and Fig.~\ref{fig:inter_calibs}. The results are also compared with the known values from the literature.

From these results, the diameters of $\epsilon$~Ant and e~Cen are in excellent agreement with those given by the JSDC when they are calibrated from one to another, and the visibilities are well fitted by an uniform disk (UD) model (see \Cref{fig:eps_ant_e_cen,fig:e_cen_eps_ant}). On the contrary, we find important inconsistencies when deriving the diameter of B~Cen or using it as a calibrator (see \Cref{fig:e_eps_ant_b_cen,fig:e_cen_b_cen,fig:b_cen_eps_ant,fig:b_cen_e_cen}). In particular this standard star is suspected to be 20$\%$ larger tan the diameter given by the JSDC ($\approx \,$3$\,$mas instead of 2.5$\,$mas). Moreover, a simple UD model seems to be unsuitable for fitting the observed visibilities (see black curve in \Cref{fig:b_cen_eps_ant,fig:b_cen_e_cen}). On the other hand the diameter of B~Cen is also well established by interferometry in the $K$ band \citep{richichi2009} and was used many times as an interferometric calibrator \citep{kervella06a,kervella2007}.  We emphasize that the transfer function was very stable during this night, thus this result cannot be attributed to the atmospheric conditions (see coherence time in Table~\ref{Tab.log}). Since the origin of this discrepancy remains unknown, this analysis prevented us for using B~Cen to calibrate the Cepheid $\ell$~Car.%

\begin{table}[h]
\caption{\label{Tab.inter_cal} \small Diameters in milliarcseconds obtained by the inter-calibration between B Cen, $\epsilon$ Ant and e~Cen. The targets of the calibration are indicated in the upper part of the table together with their angular diameter in $L$ band from the JSDC catalogue in parentheses. The first column indicates the calibrator used to obtain the derived UD angular diameters in each cell. The red cells represent discrepant values compared to the JSDC catalogue at more than 1$\,\sigma$. This deviation from the derived $\theta_\mathrm{UD}$ to the JSDC value is derived by $|\theta_\mathrm{UD} -~\theta_\mathrm{JSDC}|/\sigma_\mathrm{JSDC}$.}
\begin{center}
\begin{tabular}{l|c|c|c}
\hline
\hline
	&	e Cen	& $\epsilon$ Ant & B Cen \\
	&  (2.97$\pm$0.29)         &    (2.86$\pm$0.30)           &    (2.54$\pm$0.28)  \\
		\hline											
e Cen           & 	-    & 	2.91$\pm$0.01    & 	\cellcolor{red!25} 3.09$\pm$0.02    	\\
\hline
$\epsilon$ Ant	&2.96$\pm$0.01		& 	-    &  \cellcolor{red!25} 3.10$\pm$0.02	\\
\hline
B Cen       	&\cellcolor{red!25}	2.41$\pm$0.02	&  \cellcolor{red!25} 2.39$\pm$0.02    & 	    -	\\
\hline
\hline
\end{tabular}
\normalsize
\end{center}
\end{table}

\end{appendix}

\end{document}